\documentclass[journal]{IEEEtran}
\ifCLASSINFOpdf
\else
\fi

\usepackage{soul}
\usepackage{nomencl}
\usepackage{amsmath,amssymb}
\usepackage{float}
\usepackage{graphicx}
\usepackage{subfigure}
\usepackage{indentfirst}
\usepackage{etoolbox}
\usepackage{cite}
\usepackage{algorithmic}
\usepackage{algorithm}
\usepackage{xcolor}
\usepackage{multirow}
\usepackage{makecell}
\usepackage{booktabs}
\usepackage{dsfont}
\usepackage{textcomp}
\usepackage{threeparttable}
\usepackage{bm}
\usepackage{siunitx}
\usepackage{verbatim}
\usepackage{tabularx}
\usepackage{mathptmx} 
\usepackage{newtxtext,newtxmath}
\usepackage[T1]{fontenc}
\usepackage{textcomp}  
\usepackage{times}   
\usepackage{nomencl}
\usepackage{hyperref}

\begin{document}
\title{Two-Stage Robust Planning for Park-Level \\Integrated Energy System Considering \\ Uncertain Equipment Contingency}

\author{Zuxun Xiong,~\IEEEmembership{}
        Xinwei Shen$^*$,~\IEEEmembership{}
        Hongbin Sun~\IEEEmembership{}% <-this % stops a space\
\thanks{Z. Xiong is with the Department of Engineering Science, University of Oxford, UK (e-mail: zuxun.xiong@eng.ox.ac.uk).}
\thanks{$^*$X. Shen is with the Institute for Ocean Engineering, Tsinghua Shenzhen International Graduate School, Tsinghua University, Shenzhen 518055, China (e-mail: sxw.tbsi@sz.tsinghua.edu.cn).}% <-this % stops a space
\thanks{H. Sun is with the Department of Electrical Engineering, Tsinghua University, Beijing, China (e-mail: shb@tsinghua.edu.cn).}
\thanks{This work is partly supported by Natural Science Foundation of China under No. 52007123 (Corresponding: Xinwei Shen).}
}


\markboth{Journal of \LaTeX\ Class Files}%
{Shell \MakeLowercase{\textit{et al.}}: Bare Demo of IEEEtran.cls for IEEE Journals}



\maketitle

\begin{abstract}
%In this paper, we propose a two-stage robust planning model for park-level Integrated Energy System (IES). 
To enhance the reliability of Integrated Energy Systems (IESs) and address the research gap in reliability-based planning methods, this paper proposes a two-stage robust planning model specifically for park-level IESs. The proposed planning model considers uncertainties like load demand fluctuations and equipment contingencies, and provides a reliable scheme of equipment selection and sizing for IES investors. Inspired by the unit commitment problem, we formulate an equipment contingency uncertainty set to accurately describe the potential equipment contingencies which happen and can be repaired within a day. Then, a modified nested column-and-constraint generation algorithm is applied to solve this two-stage robust planning model with integer recourse efficiently. In the case study, the role of energy storage system for IES reliability enhancement is analyzed in detail. Computational results demonstrate the advantage of the proposed model over other planning models in terms of improving reliability. 
\end{abstract}

\begin{IEEEkeywords}
Park-level Integrated Energy System, reliability, robust planning model, energy storage system, nested column-and-constraint generation.
\end{IEEEkeywords}

% \fi
\IEEEpeerreviewmaketitle

\section*{Nomenclature}

% Group A
\subsection*{A. Indices and Sets}
\begin{tabular}{p{1.5cm}p{6.1cm}}  
$\mathbb{X}$ & Set of investment decision variable $X$. \\
$\mathbb{L}$ & Set of uncertain loads. \\
$\mathbb{S}$ & Set of operational state of components. \\
0,1 & Indices for normal operation scenario and scenario with contingency. \\
$n,i,j,k$ & Indices for different equipment options. \\
$t$ & Index for time slot. \\
$d,e,h,c,g$ & Indices for different energies. \\
$yr$ & Index for year. \\
B,T & Indices for BESS and TESS. \\
$ch/dis$ & Indices for charge and discharge. \\
$f$ & Index for feeders from substation. \\
$q$ & Index for iteration in outer algorithm. \\
$r$ & Index for iteration in inner algorithm. \\
\end{tabular}

% Group B
\subsection*{B. Parameters}

\begin{tabular}{p{1.5cm}p{6.3cm}}  
$IC^{\text{Equi}}_{n}$ & Investment cost for equipment $n$. \\
$IC^{\text{ESS}}_E,IC^{\text{ESS}}_P$ & Investment cost for ESS energy/power capacity. \\
$m$ & Discount factor for total cost. \\
$\mathcal{R}^{e}_{t},\mathcal{R}^{g}_{t}$ & Time-of-use price for electricity and gas. \\
$p^{\text{LS},d}_{t}$ & Penalty cost for shedding load $d$ at time $t$. \\
$PP$ & Total year of the planning period. \\
$\gamma$ & Discount rate. \\
$C^{\text{SUB}}_{e-e}$ & Energy conversion rate of substation. \\
$P^{\text{SUB}}_{f,max}$ & Maximum power from substation through feeder $f$. \\
$C^{\text{Equi}}_{a-b}$ & Conversion rate of Equi from energy a to b. \\
$P^{\text{Equi}}_{n,max}$ & Maximum output power of Equi $n$. \\
$T^{ch}_{max}$ & Maximum charging cycles of the battery. \\
$\text{SoC}^{\text{B/T}}_{min/max}$ & Min/Maximum state of charge of B/TESS. \\
$\eta^{\text{B/T}}_{ch/dis}$ & Charging/discharging efficiency of B/TESS. \\
Period & The length of planning period. \\
$N^{\text{Equi}/f}$ & The number of options of Equi/feeders. \\
$\Gamma^{N},\Gamma^{I/D}_{n/f}$ & Number of simultaneous component failure, failure interval, failure duration. \\
$\Gamma^{l}$ & Maximum load fluctuation. \\
$\Delta l^{d}_t$ & Fluctuation value of load $d$ at time $t$ \\
\end{tabular}

% Group C
\subsection*{C. Variables}

\begin{tabular}{p{1.5cm}p{6.3cm}}  
$X^{\text{Equi}}_n$ & Investment state of equipment $n$. \\
$X^{\text{ESS}}_E,X^{\text{ESS}}_P$ & Investment energy/power capacity of ESS. \\
$P^{\text{SUB}}_{0/1,t}$ & Output power of substation at time $t$ in scenario 0/1. \\
$P^{\text{Equi}}_{0/1,n,t}$ & Output power of Equipment $n$ at time $t$ in scenario 0/1. \\
$P^{\text{LS,d}}_{1,t}$ & Load shedding of type $d$ at time $t$. \\
$P^{\text{B/T}}_{0/1,ch/dis,t}$ & Charging/Discharging power of B/TESS at time $t$ in scenario 0/1. \\
$E^{\text{B/T}}_{0/1,t}$ &  Energy of B/TESS at time $t$ in scenario 0/1. \\
$Z^{\text{B/T}}_{0/1,ch,t}$ & Charging state of B/TESS at time $t$ in scenario 0/1. \\
$s^{\text{Equi}}_{n,t},s^{\text{Feed}}_{f,t}$ & Operational state of equipment $n$/feeder $f$ at time $t$. \\
$y^{\text{Equi}}_{n,t},y^{\text{Feed}}_{f,t}$ & Failure starting point of $n$/$f$ at time $t$. \\
$l^d_t$ & Uncertain load $d$ at time $t$. \\
$u^{d}_{t,+}/u^{d}_{t,-}$ & Up/downward fluctuations of load $d$ at $t$. \\
$\pi^1_{t,r},\ldots,\pi^{29}_{t,r}$ & Dual variables from strong duality theory. \\
$\alpha_{0,t,r},\rho_{0,t,r}$ & Dual variables from KKT conditions. \\
\end{tabular}

\section{Introduction}

\IEEEPARstart{T}o address the challenges posed by low energy utilization efficiency in traditional energy systems, Integrated Energy Systems (IESs) have been developed rapidly in recent years \cite{RN89}. Among these, developing regional IESs that directly connect energy generation and consumption is considered a promising solution \cite{RN287}, \cite{RN50}. These types of IESs, referred to as park-level IESs and represented by industrial parks, hospitals, and residential complexes, bear a substantial and gradually increasing portion of the overall societal load demand \cite{RN430}. Therefore, there is an increasing need for research into new methods that can help plan and operate such systems. However, existing studies mainly focus on the economic operation of IESs, such as day-ahead scheduling \cite{RN49}, demand response \cite{RN52}, and benefit allocation \cite{RN50phd}, with little attention given to their planning problems that can guide investment. Moreover, although reliability is critical in the aforementioned typical application scenarios of IESs, few studies emphasize this criterion. In this paper, we focus on park-level IES planning problem, where we seek to identify optimal long-term investment decisions. % that appropriately hedge against IES uncertainties.
In addition to economic efficiency, reliability is a main focus of this study. Our goal is to invest an IES which can be operated normally under uncertain perturbations, namely a robust IES, at the minimum cost.

First of all, relevant indices and methods should be introduced to assess the reliability of the IES. There is a comprehensive reliability assessment framework for the traditional power system. Bilinton et al. define the reliability of power system and summarize the reliability indices to reflect different characteristics \cite{RN200}. Indices like loss of load frequency (LOLF), loss of load expectation (LOLE) and expected energy not supplied (EENS) are widely used to evaluate the reliability of power systems. They can also be applied to evaluate IES reliability since they can reflect the operating status of a system  regardless of the energy type. The reliability assessment techniques fall into two categories: simulation and analytical methods. Simulation approaches, e.g., Monte Carlo Simulation (MCS), estimate reliability indices by simulating the actual and random processes of system \cite{RN286}, while analytical methods use models to mathematically calculate these indices \cite{RN199}. Both techniques are applicable for evaluating IES reliability \cite{RN210}.

%R-based planning
% Although the reliability assessment method has been widely studied, this process is a posteriori, i.e., the assessment result is given only after the construction plan and operation scheme of the energy system are known. Even though the assessment result could provide a reference for the reconfiguration of the energy system, the reconfiguration process will cost extra and waste a lot of time. Therefore, researchers turn to focus on how to integrate reliability indices with the planning model of the IES, namely reliability-based planning. The aim of these studies is to provide a planning scheme that will enable the IES to operate normally despite potential uncertain interference. We divide these studies into two categories according to the sources of uncertainty considered. 
However, the reliability assessment is \textit{a posteriori}, i.e., the assessment result is given only after the construction plan and operation scheme of the energy system are known. To address this limitation, researchers try to integrate these reliability indices into the planning models, leading to reliability-based planning approaches. The aim of these studies is to provide a planning scheme that will enable the energy systems to operate normally despite potential uncertain interference. Since system reliability is primarily affected by two types of uncertainties, reliability-based planning specifically targets these factors.

%考虑external uncertainty
The first group consists of external factors such as variations in renewable resource output \cite{RN361} and fluctuating load demands. Reference \cite{RN295} proposes a two-stage robust planning model, considering random wind output and uncertain loads. The numerical cases show that the planning results can make the system immune to the fluctuations of the considered uncertainties, and no more load shedding occurs, thus improving the system reliability. Stefano et al. propose a robust optimization framework for long-term energy system planning that considers various external uncertainties \cite{RN309} and apply it to plan the real-world European power system \cite{RN308}. 
%和contingency有关的r-based planning 
    %加入RA的r-based planning
While external uncertainties affect system reliability, the internal uncertainties such as the failure of critical components like transmission lines and generators, have a more immediate and substantial impact on system reliability and economy \cite{RN360}. Consequently, an increasing number of reliability-based planning methods place greater emphasis on addressing the challenges posed by component contingencies. A common and simple approach is introducing security constraints like N-1 security criterion, which is widely used in power system \cite{RN356}. Reference \cite{RN171} incorporates the N-1 security criterion into a planning model for coupled electricity-gas transmission systems. Our previous study also proposes an N-1 planning model for IESs \cite{RN104}. This criterion is extended to an N-k-$\epsilon$ criterion, and integrated into planning models \cite{RN220}, \cite{RN358}.

%MCS
    %analytical
Due to the over conservatism of the N-1 criterion, some studies integrate the aforementioned reliability assessment process into the planning model and use reliability metrics as constraints to ensure system stability. 
% Some studies formulate the security constraint by integrating the reliability assessment process into the planning model. As mentioned before, the reliability assessment process requires known planning results; therefore, most of the methods that integrate reliability assessment into planning models apply decomposition schemes to solve the model. That is, after obtaining a planning result, the reliability of the temporary planning scheme will be evaluated in a subproblem, and then relevant constraints or variables will be added back to the primal problem. The most reliable planning result will be obtained through iterations.    
The first group of methods incorporates simulation-based reliability assessment process into planning model \cite{RN218,RN217,RN170,RN147,RN179,RN355,RN359}. To be specific, \cite{RN218} generates numerous scenarios to simulate microgrid operation, considering the component outages and uncertain renewable resources output to calculate and limit the LOLE. A tri-level power system expansion planning framework is proposed by \cite{RN217}. Its first two levels correspond to construction and operation decisions respectively, and the third level applies MCS for reliability assessment. 
% If the reliability index does not meet the criteria, the reliability subproblem of the third level will add the reliability dual cut back to the first level. Finally, the solution converges and provides the optimal planning scheme. 
A similar three-stage planning model for electricity and gas systems is proposed by \cite{RN170}. \cite{RN147} refers to the risk measure Value-at-Risk as reliability index and calculates it in the subproblem using MCS. \cite{RN179} proposes a bi-level optimal ESS planning model for IES, which generates scenarios based on a two-status model for component, and applies MCS method to calculate the EENS of the system. In \cite{RN355}, the fuzzy set theory is combined with MCS to efficiently calculate the reliability indices, which are then tested by the reliability constraints in the subproblem. \cite{RN359} generates several groups of scenarios and obtains a planning scheme pool based on them. Then their reliability indices are calculated and compared.

Since these methods are iterative optimization-simulation approaches, their results are typically suboptimal. Besides, reliability simulation under a large number of scenarios imposes an ultra-high computational burden \cite{RN351}. Consequently, planning methods utilizing analytical approaches have attracted growing attention in recent years.
Reference \cite{RN197} uses analytical approach to evaluate the LOLE of the planning scheme for electricity and gas system, then constructs relevant constraints and adds them back to the planning model. Based on graph theory, \cite{RN148} proposes a distribution network expansion planning model which explicitly incorporates analytical reliability assessment. 
\cite{RN353} also takes advantage of radial operation of distribution networks and formulates a planning model with reliability indices like EENS as a mixed-integer linear programming (MILP). However, most existing analytical-based methods focus on distribution network expansion and are not applicable to scenarios involving generation outages \cite{RN357}.
    %RO modeling
%Most of these studies that consider possible component contingencies also consider the effects of external uncertainties. Robust optimization is the first choice for these studies to handle and model the uncertainties. As mentioned earlier, robust models find a possible worst-case scenario under uncertainty and optimize the worst-case scenario. However, these studies also differ in the target when applying robust optimization. For example, some studies formulate tri-level planning models where they find the worst-case scenario under uncertain load and renewable resources output in the second level, and then optimize the planning result \cite{RN217,RN170}. Meanwhile, other two studies use robust optimization models to find the worst-case scenarios under uncertain device contingencies. An N-k-$\epsilon$ planning model based on the N-1 criterion is proposed by \cite{RN220} to ensure that the system can still meet the $1-\epsilon$ ratio of the total load demand in the presence of k simultaneous contingencies. When k contingencies are considered simultaneously, the combination explosion will impose a large computational burden, so this study uses a robust optimization approach to find the worst-case under contingencies first, thus improving the solution efficiency. A tri-level planner-attacker-defender robust model is proposed by \cite{RN251}, where the planner gets the planning scheme in the first level, the attacker damages the components in the second level to reach the worst case, and the defender optimizes the worst case again in the third level.

%gap
Most of these studies focus on power systems. As mentioned earlier, there is an urgent need for a reliability-based planning method that can effectively guide the development of IESs. Moreover, while power system reliability-based planning methods offer useful insights, they have notable limitations. First, many studies fail to adequately consider both external and internal uncertainties. The few studies that do consider both types of uncertainties consider them separately in different subproblems, overlooking their synergies, which can potentially lead to suboptimality. Additionally, the modeling of component contingencies is often inappropriate. For example, security criterion assumes the component is out of operation during the whole planning period instead of a specific period \cite{RN356,RN171,RN104,RN220,RN358,RN170,RN179}. Even in recent analytical-based approaches, real reliability parameters, such as failure and repair rates, are not fully integrated into the models. Instead, reliability is calculated based on components being entirely out of operation, using specific network structures. This will inevitably lead to a discrepancy between the model and reality.

To address these problems, this paper proposes a two-stage robust planning model for park-level IES. Both external uncertainties and equipment contingencies are considered. An uncertainty set is proposed to accurately model the equipment contingencies. Furthermore, we develop a nested algorithm to solve the model. Simulation is carried out in a real-world industrial park to verify the effectiveness of the proposed model and algorithm. The main contributions are:

\begin{itemize}
  \item[1)] We propose a two-stage robust planning model to make a tradeoff between economy and reliability for the park-level IES. The first stage optimizes the planning scheme, including equipment selection and sizing, while the second stage determines the optimal operation, such as generator and ESS dispatch, to against the worst-case scenario identified by the uncertainty sets.
  % The worst-case scenario here is determined by a combination of external uncertainties and equipment contingencies.

  \item[2)] We propose an uncertainty set for component that corresponds to real reliability parameters, accurately modeling  equipment contingencies within the two-stage robust planning model.
  
  \item[3)] We adapt a nested column-and-constraint generation (C\&CG) algorithm to solve the robust planning model and proposed an exact linearization method for the strong duality-based nested C\&CG algorithm. Its computational efficiency is validated through case studies.

  % \item[4)] Based on the numerical case, we analyze the effect of different budgets of the proposed uncertainty set on the planning results in detail. The role of ESS in improving IES reliability is also analyzed.
\end{itemize}

The rest of the paper is organized as follows. Section II formulates the two-stage robust IES planning model based on the Energy Hub (EH) model. Section III introduces the solving algorithms. Section IV presents the results of the case study and analyzes them in detail. Section V further discusses our algorithm. The conclusion drawn from the case study is provided in Section VI.

\section{Two-Stage Robust IES Planning Model}
 % We formulate the two-stage robust planning model in this section. First, we introduce the EH model and construct the planning model based on it. Then, we explain the proposed uncertainty set in detail. 

\subsection{Planning Model Formulation}
% As we mentioned, the main object of this study is the park-level IES. Due to its relatively small scale, it is unnecessary to consider its energy transmission network. 
We use the EH model, which reflects the relationship between energy input and output \cite{RN93}, for park-level IES modeling. We follow the planning model of our previous study \cite{RN104}, where three forms of energy are considered, i.e. electricity, heat, and cooling. The model makes decisions regarding the installation of combined cooling, heat and power (CCHP) units, gas boilers (GB), and electric chillers (EC), and can also choose to import power from a substation (SUB) through the distribution lines connected to the IES. We also introduce multi-energy ESS, e.g., battery ESS (BESS) and thermal ESS (TESS), as options in this model. We will use B and T as abbreviation for BESS and TESS later.  
%Among them, the CCHP unit generates three different kinds of energies by consuming natural gas. The gas boiler provides heat also by consuming natural gas, and the electric chiller converts electricity into cooling energy. 
%In addition, there is a substation (SUB) to provide extra electricity supply to this regional IES. 
Based on the EH hub and the proposed equipment contingency set, we formulate the two-stage robust planning model for IES as follows.

\subsubsection{Objective Function} The objective function minimizes the total cost. To reflect the impact of uncertainties and ensure robustness, we formulate it in the following two-stage structure:
\begin{subequations}
    \begin{align}
 \min_{X\in{\mathbb{X}}} & f^{\text{inv}}+\max_{l\in{\mathbb{L}},s\in{\mathbb{S}}}\min_{(Z,P)\in{F(x,l,s)}}f^{\text{ope}}_0+f^{\text{shed}}_1,
    \label{eq:3-ro-obj}\\
f^{\text{inv}}&=\sum_{\text{Equi}}\sum_{n\in\{i,j,k\}}IC^{\text{Equi}}_{n} X^{\text{Equi}}_n+\sum_{\text{ESS}}IC^{\text{ESS}}_E X^{\text{ESS}}_E +IC^{\text{ESS}}_P X^{\text{ESS}}_P,
    \label{eq:2-obj-inv}\\
    f^{\text{ope}}_0 &= m\sum_{t}[P^{\text{SUB}}_{0,t}\mathcal{R}^{e}_{t}+(\sum_i P^{\text{CCHP}}_{0,i,t}+\sum_j P^{\text{GB}}_{0,j,t})\mathcal{R}^{g}_{t}],
    \label{eq:2-obj-ope}\\
    f^{\text{shed}}_1 &= m\sum_{t} \sum_{d\in\{e,h,c\}} P^{\text{LS},d}_{1,t}p^{\text{LS},d}_{t},\label{eq:2-obj-shed}\\
    \text{Equi} &\in \{{\text{CCHP},\text{GB},\text{EC}}\}, \quad \text{ESS} \in \{{\text{B},\text{T}}\}.\notag
\end{align}
\end{subequations}

%the investment cost, operational cost, and penalty for shedding loads are denoted by $f^{inv}$, $f^{ope}_0$ and $f^{shed}_1$, respectively. 
Here, the objective function \eqref{eq:3-ro-obj} minimizes the total investment cost, defined by the function $f^{\text{inv}}$, as well as the worst-case operational cost across all possible load and equipment failure scenarios, denoted by $\mathbb{L}$ and $\mathbb{S}$. This worst-case cost is defined by the max-min function which considers both the operational cost and load shed penalty, defined by functions $f^{\text{ope}}_0$ and $f^{\text{shed}}_1$, respectively.  
%The variable $\mathbf{X}$ is the vector of the first-stage investment decision variables which decide the construction scheme of equipment, and give rise to the expression for the investment cost in \eqref{eq:2-obj-inv}. The investment decisions $\mathbf{X}$ for CCHP, GB, and EC are binary variables ($X=1$ if the equipment is being built and $\mathbf{X}$ if not), while $X$ is a continuous variable ESS, representing the capacity of ESS to be built. It considers all possible scenarios under uncertainty sets and leads to a robust solution. 
The sets $\mathbb{L}$ and $\mathbb{S}$ are the uncertainty sets for loads and equipment operating status. Continuous variables $P$ and binary variables $Z$ are second-stage operational decisions which represent the output of equipment and the storage status of ESS ($Z=1$ for charging and $Z=0$ for discharging) respectively. The feasible set of the second stage, represented by $F(\text{•}),$ is a function of $X$, $l$, and $s$. Note that the second-stage variables $P$ and $Z$ are fully adaptive to any realization of the uncertainties \cite{RN150}.

Equation \eqref{eq:2-obj-inv} shows that the investment cost includes the investment of all kinds of optional equipment and ESS. Here $\text{Equi}\in\{\text{CCHP}, \text{GB}, \text{EC}\}$, $n$ could be $i, j, k$ when $\text{Equi}$ stands for different type of equipment. The binary variable $X^{\text{Equi}}_n$ is the first-stage investment decision variable which decides the construction scheme of the $n^{\text{th}}$ equipment ($X=1$ if the equipment is being built and $X=0$ if not), and give rise to the expression for the investment cost in \eqref{eq:2-obj-inv}. The investment decision variables for ESS include energy capacity $X^{\text{ESS}}_E$ and power capacity $X^{\text{ESS}}_P$, where $\text{ESS}\in{\text{B}, \text{T}}$. These continuous variables represent the storage capacity and maximum charging/discharging power of ESS. The unit investment cost is given by $IC^{\text{Equi}}_n$ for three types of equipment, and by $IC^{\text{ESS}}_E$ and $IC^{\text{ESS}}_P$ for ESS. %It considers all possible scenarios under uncertainty sets and leads to a robust solution. 

Equation \eqref{eq:2-obj-ope} is the operational cost, which reflects the cost of procuring electricity and natural gas. We use subscripts 0 and 1 to denote normal operation scenario and scenario with contingencies, respectively. $P^{\text{Equi}}_{0,n,t}$ represents the power output of equipment $n$ at time $t$ in normal operation scenario. Besides three kinds of equipment to be invested, we also consider an external substation which supplies power to the IES through distribution lines and use SUB to represent it. So, $P^{\text{SUB}}_{0,t}$ represent the power output of the substation at time $t$ in normal operation scenario.
$\mathcal{R}^e_t$ and $\mathcal{R}^g_t$ stand for time-of-use price of electricity and natural gas respectively. In \eqref{eq:2-obj-shed}, $P^{\text{LS},d}_{1,t}$ with $d\in\{e,h,c\}$ represents the electricity, heating or cooling loads shedding, respectively, and $p^{\text{LS},d}$ is the corresponding penalty cost for each type of load. 
% Since the occurrence of equipment outage is an event with low probability, the total operating cost will be calculated using the variables from the normal operation scenario, $f^{ope}_0$, while the penalty for load shedding will be calculated using variables from scenario with contingency, $f^{shed}_1$. 
We use $m$ to convert all the operational costs in the planning period into the planning year through the discount rate $\gamma$:
$
m=\sum_{yr=1}^{P P} \frac{365}{(1+\gamma)^{yr-1}},
$
where $PP$ represents the total years of the planning period.

\subsubsection{Constraints}As we mentioned, we divide variables into two sets. For normal operation scenario, the constraints can be expressed as:
% L<=CP
\begin{subequations}
    \begin{align}
    &\textbf{L} \leq \textbf{C}^{\text{SUB}}\textbf{P}^{\text{SUB}}_0+\sum_{\text{Equi}}\textbf{C}^{\text{Equi}}\textbf{P}^{\text{Equi}}_0-\textbf{P}^{\text{ESS}}_0, \label{eq:con-balance-0}\\
    &l^e_t \leq C^{\text{SUB}}_{e-e}P^{\text{SUB}}_{0,t}+\sum_{i}C^{\text{CCHP}}_{i,g-e}P^{\text{CCHP}}_{0,i,t}+\sum_{k}C^{\text{EC}}_{k,e-e}P^{\text{EC}}_{0,k,t}-\notag \\
    &P^{\text{B}}_{0,ch,t}+P^{\text{B}}_{0,dis,t}, \label{eq:con-balance-e-0}\\
    &l^h_t \leq \sum_{i}C^{\text{CCHP}}_{i,g-h}P^{\text{CCHP}}_{0,i,t}+ \sum_{j}C^{\text{GB}}_{j,g-h}P^{\text{GB}}_{0,j,t}-P^{\text{T}}_{0,ch,t}+P^{\text{T}}_{0,dis,t},\\
    &l^c_t \leq \sum_{i}C^{\text{CCHP}}_{i,g-c}P^{\text{CCHP}}_{0,i,t}+\sum_{k}C^{\text{EC}}_{k,e-c}P^{\text{EC}}_{0,k,t}, \label{eq:con-balance-c-0}\\
%upper and lower output
    &0 \leq P^{\text{Equi}}_{0,n,t} \leq P^{\text{Equi}}_{n,max}X^{\text{Equi}}_n, \label{eq:con-output-0}\\
        &0 \leq P^{\text{SUB}}_{0,t} \leq \sum_f^{N^f} P^{\text{SUB}}_{f,max}, \label{eq:con-sub}\\
%BESS
    &\sum_t P^{\text{B}}_{0,ch,t}-T^{ch}_{max}X^{\text{B}}_E \leq 0, \label{eq:con-max-cycle}\\
    &\text{SoC}^{\text{B/T}}_{min}X^{\text{B/T}} \leq E^{\text{B/T}}_{0,t} \leq \text{SoC}^{\text{B/T}}_{max}X^{\text{B/T}}, \label{eq:con-SoC}\\
%discharge and charge power limits
    &0 \leq P^{\text{B/T}}_{0,ch,t} \leq MZ^{\text{B/T}}_{0,ch,t}, \label{eq:con-chdis-1}\\
    &0 \leq P^{\text{B/T}}_{0,dis,t} \leq M(1-Z^{\text{B/T}}_{0,ch,t}), \label{eq:con-chdis-2}\\
    &P^{\text{B/T}}_{0,ch/dis,t} \leq X^{\text{B/T}}_P, \label{eq:con-chdis-3}\\
%Energy of ESS
    &E^{\text{B/T}}_{0,t} = E^{\text{B/T}}_{0,t-1}+P^{\text{B/T}}_{0,ch,t-1}\eta^{\text{B/T}}_{ch}-P^{\text{B/T}}_{0,dis,t-1}/\eta^{\text{B/T}}_{dis}, \label{eq:con-Energy}\\
    &X \in{\mathbb{X}}= \left\{
    \begin{gathered}
        X^{\text{ESS}}_E \geq X^{\text{ESS}}_P \geq 0, 
        X^{\text{Equi}}_n \in \{0,1\} \label{eq:set-x}
    \end{gathered}\right
    \}, \\
    &Z^{\text{B/T}}_{ch,t} \in \{0,1\}, l \in{\mathbb{L}},  s \in{\mathbb{S}}, \label{eq:set-z}
\end{align}
\end{subequations}
\begin{equation}
    \begin{gathered}
        \forall 1 \leq t \leq \text{Period}, \\
        \forall 1 \leq i \leq N^{\text{CCHP}}\text{,} \quad
        \forall 1 \leq j \leq N^{\text{GB}}\text{,} \quad
        \forall 1 \leq k \leq N^{\text{EC}}. \notag
    \end{gathered}
\end{equation}

Here $\text{Period}$ is the length of the planning period, $N^{\text{Equi}}$ represents the number of optional equipment of a certain type. Equation \eqref{eq:con-balance-0} is the supply-demand balance constraint under normal operation scenario formulated based on EH model. where $\mathbf{L}$ is the matrix form of $l^d_t$, which denotes the loads of IES. %$e$, $h$ and $c$ represent electricity, heat, and cooling respectively.
$\mathbf{P}^{\text{Equi}}_0$ is the matrix form of $P^{\text{Equi}}_{0,n,t}$, which denotes the output of different equipment, $\mathbf{C}^{\text{Equi}}$ represents the coupling matrix of different equipment and the subscripts indicate different kinds of energy. For example, $C^{\text{CCHP}}_{i,g-e}$ represents the efficiency of the $i^{\text{th}}$ CCHP in converting natural gas into electricity. All supply must be equal or larger than the demand at any time under normal operation scenario (no shedding load is allowed). $\mathbf{P}^{\text{ESS}}$ denotes the matrix of charging/discharging power of BESS and TESS. The subscript $ch$ and $dis$ denote charge and discharge, respectively. Constraint \eqref{eq:con-balance-0} can be further expressed as \eqref{eq:con-balance-e-0}—\eqref{eq:con-balance-c-0}. As indicated by constraint \eqref{eq:con-output-0}, the output of each equipment is bounded by both its lower/upper limit and its investment state $X^{\text{Equi}}_n$. Constraint \eqref{eq:con-sub} means the maximum power supplied from the external substation is the sum of the maximum line capacities of all feeders connected to the IES. $N^f$ is the number of feeders. $T^{ch}_{max}$ represents the maximum charging cycle of the BESS. Constraint \eqref{eq:con-max-cycle} limits the maximum charge amount. 
Constraint \eqref{eq:con-SoC} limits the state of charge (SoC) level of ESS between a certain range, where $E^{\text{ESS}}_{t}$ represents the stored energy of ESS at time $t$. Constraints \eqref{eq:con-chdis-1} and \eqref{eq:con-chdis-2} guarantee that there will be no simultaneous charging and discharging by using a large constant $M$. Constraint \eqref{eq:con-chdis-3} limits the charging and discharging power rate of ESS by the invested power capacity. Equation \eqref{eq:con-Energy} represents the energy variation of the ESS during the whole period. \eqref{eq:set-x} is the set for investment decision variable $X$, as we mentioned before, $X^{\text{ESS}}_E$ and $X^{\text{ESS}}_P$ for ESS capacity are continuous variables and $X^{\text{Equi}}$ for all $\text{Equi}$ are binary variables. \eqref{eq:set-z} shows that the variable $Z^{\text{B/T}}_{ch,t}$ for the ESS charge state is binary. Load $l$ and component operational state $s$ belong to two uncertainty sets we propose respectively.
We denote constraints for ESS under normal operation scenario, i.e., \eqref{eq:con-max-cycle}—\eqref{eq:con-Energy}, as:
\begin{equation}
Cons^{\text{ESS}}_0(X^{\text{B/T}}_E, X^{\text{B/T}}_P,P^{\text{B/T}}_{0,ch/dis,t},Z^{\text{B/T}}_{0,ch,t},E^{\text{B/T}}_{0,t}). \label{eq:con-ESS-0}
\end{equation}

For constraints under scenario with contingencies, we add load shedding term to the supply-demand balance constraint to guarantee the feasibility of the problem:
\begin{equation}
    \textbf{L} \leq \textbf{C}^{\text{SUB}}\textbf{P}^{\text{SUB}}_1+ \sum_{\text{Equi}}\textbf{C}^{\text{Equi}}\textbf{P}^{\text{Equi}}_1-\textbf{P}^{\text{ESS}}_1+\textbf{P}^{\text{LS}}_1 .\label{eq:con-balance-1}
\end{equation}
Constraints on shedding load $P^{\text{LS},d}_{1,t} \ge 0$ should also be considered. Then, the upper and lower limits of equipment output are constrained by both the investment decision variable $X$ and the operation state variable $s$,
\begin{equation}
    0 \leq P^{\text{Equi}}_{1,n,t} \leq P^{\text{Equi}}_{n,max}s^{\text{Equi}}_{n,t}X^{\text{Equi}}_n \label{eq:con-output-1},
\end{equation}
which is different from that of the normal operation scenario as shown in constraint (\ref{eq:con-output-0}). Constraint (\ref{eq:con-output-1}) indicates that even if the equipment is invested, its output power will be 0 when it fails ($s=0$). Also, for the substation, we consider the contingency of each feeder by introducing operation state $s^{\text{feed}}_{f,t}$ for it. Then we have the output range of substation in scenario with contingencies as:
\begin{equation}
    0 \leq P^{\text{SUB}}_{1,t} \leq \sum_f P^{\text{SUB}}_{f,max}s^{\text{feed}}_{f,t}.
\end{equation}
The ESS-related constraints under scenario with contingency are almost the same as those under normal operation scenario, except for the removal of the lower and upper limits of the SoC. This means when contingencies occur, the ESS can be fully charged and discharged. We denote these ESS-related constraints as follow for convenience:
\begin{equation}
Cons^{\text{ESS}}_1(X^{\text{B/T}}_E,X^{\text{B/T}}_P,P^{\text{B/T}}_{1,ch/dis,t},Z^{\text{B/T}}_{1,ch,t},E^{\text{B/T}}_{1,t}).  \label{eq:con-ESS-1}
\end{equation}

\subsection{Uncertainty Set Formulation}
In this part, we propose a unique uncertainty set to describe component contingencies inspired by typical Unit Commitment (UC) problem \cite{RN150}. UC problem involves determining which and when generating units should be turned on and turned off over a certain period of time, typically ranging from several hours to a few days. There are resemblances between the modeling of unit on/off states in UC problem and the operating statuses of equipment in our problem. For example, the constraints associated with turn-on and turn-off actions of the UC problem can be used to describe the start point and end point of an equipment contingency. The minimum up and minimum down times constraints of the UC problem can be used to describe the duration of an equipment contingency. With this insight, we formulate the following uncertainty set for component contingencies:
\begin{equation}
    {\mathbb{S}}= \left\{ s^{\text{Equi}}_{n,t},s^{\text{Feed}}_{f,t} \, \left\vert 
    \begin{gathered}
    \sum_{\text{Equi}}\sum_{n}s^{\text{Equi}}_{n,t}+\sum_{f}s^{\text{Feed}}_{f,t} \geq \sum_{\text{Equi}}N^{\text{Equi}}+N^f-\Gamma^{N} \\
        \sum_{\text{Equi}}\sum^{t\in [\Gamma^{I}_{n/f},\text{Period}]}_{v=t-\Gamma^{I}_{n/f}+1}y^{\text{Equi/Feed}}_{n/f,v} \leq 1 \\
        \sum^{t\in [\Gamma^{D}_{n/f},\text{Period}]}_{v=t-\Gamma^{D}_{n/f}+1}y^{\text{Equi/Feed}}_{n/f,v} = 1-s^{\text{Equi/Feed}}_{n/f,t} \\
        y^{\text{Equi}}_{n,t} \in \{0,1\}, s^{\text{Equi}}_{n,t},s^{\text{Feed}}_{f,t}\in \{0,1\} \notag
    \end{gathered} \right. 
    \right\}
\end{equation}

where $s^{\text{Equi}}_{n,t}$ and $s^{\text{Feed}}_{f,t}$ represent the operation state of equipment $n$ and feeder $f$ at time $t$, respectively. $y^{\text{Equi}}_{n,v}$ and $y^{\text{Feed}}_{f,v}$ are binary variables indicating the starting point of a certain contingency of the equipment $n$ or feeder $f$. All $\Gamma$ are the constants used to control the conservativeness of the model, which we also call budgets. For example, $\Gamma^{N}$ is a budget for the number of simultaneously failed equipment and feeder. When it equals to 2, the first constraint of the uncertain set $\mathbb{S}$ means that up to 2 equipment and feeder are out of operation at the time $t$. $\Gamma^{I}_{n/f}$ and $\Gamma^{D}_{n/f}$ denote the interval between two sequential failures of equipment $n$ or feeder $f$, and the failure duration of $n$ or $f$ respectively. So, the second and third constraints are used to describe the interval and the duration of contingencies. As for the uncertainty set $\mathbb{L}$ for multi-energy loads, we formulate a similar one as that in \cite{RN217}:
\begin{equation}
    {\mathbb{L}}= \left\{ l^d_t \in \mathbb{R}_{\geq 0} \, \left\lvert 
    \begin{gathered}
        l^{d}_t = \bar{l}^{d}_t+\Delta l^{d}_t u^d_{t,+} -\Delta l^{d}_tu^d_{t,-}   \\
        \sum^{\text{Period}}_{t=1}(u^d_{t,+}+u^d_{t,-}) \leq \Gamma^{l} \\ 
        u^d_{t,+}+u^d_{t,-} \leq 1, \quad u^d_{t,+}, u^d_{t,-} \in \{0,1\} \notag
    \end{gathered} 
    \right.
    \right\}.
\end{equation}
Here $\bar{l}^{d}_t$ is the nominal load demand at time $t$. $d$ can represent three different kinds of energy. $\Delta l^{d}_t$ is the fluctuation value of the load at time $t$. We use binary variables $u^d_{t,+}$ and $u^d_{t,-}$ to represent upward and downward fluctuations of load, respectively. $\Gamma^l$ is the budget of this uncertainty set, limiting the maximum fluctuation, i.e., the sum of the binary variables $u^d_{t,+/-}$, throughout the entire planning period.

\section{Solving Method}

\subsection{Two-Stage Robust Planning Model Reformulation}
 The widely-used C\&CG algorithm \cite{RN235} for solving  two-stage robust optimization problems cannot be used to solve our planning model due to the existence of the integer recourse (the second-stage variable). Therefore, we turn to nested C\&CG algorithm \cite{RN303}. Firstly, the proposed two-stage robust planning model should be reformulated into a master problem and a subproblem. Naturally, the master problem will be:
\begin{subequations}
    \begin{align}
&\textbf{MP:   } \\
&\min_{X\in{\mathbb{X}}}\sum_{\text{Equi}}\sum_{n}IC^{\text{Equi}}_{n} X^{\text{Equi}}_n+\sum_{\text{ESS}}IC^{\text{ESS}}_E X^{\text{ESS}}_E +IC^{\text{ESS}}_P X^{\text{ESS}}_P+\psi , \\
& s.t. \quad\psi \geq m\sum_{t}[P^{\text{SUB}}_{0,t,q}\mathcal{R}^{e}_{t}+
(\sum_i P^{\text{CCHP}}_{0,i,t,q}+
\sum_j P^{\text{GB}}_{0,j,t,q})\mathcal{R}^{g}_{t}
+\notag \\ 
&P^{\text{LS},d}_{1,t,q}p^{\text{LS},d}_{t}] ,\label{<Cons_MP_begin>}\\
    &\widetilde{\textbf{L}} \leq \textbf{C}^{\text{SUB}}\textbf{P}^{\text{SUB}}_{0,q}+\sum_{\text{Equi}}\textbf{C}^{\text{Equi}}\textbf{P}^{\text{Equi}}_{0,q}-\textbf{P}^{\text{ESS}}_{0,q}, \\
    &\widetilde{\textbf{L}} \leq \textbf{C}^{\text{SUB}}\textbf{P}^{\text{SUB}}_{1,q}+\sum_{\text{Equi}}\textbf{C}^{\text{Equi}}\textbf{P}^{\text{Equi}}_{1,q}-\textbf{P}^{\text{ESS}}_{1,q}+\textbf{P}^{LS}_{1,q}, \\
    &0 \leq P^{\text{Equi}}_{0,n,t,q} \leq P^{\text{Equi}}_{n,max}X^{\text{Equi}}_{n},\\
    &0 \leq P^{\text{Equi}}_{1,n,t,q} \leq P^{\text{Equi}}_{n,max}\Tilde{s}^{\text{Equi}}_{n,t}X^{\text{Equi}}_{n},\\
    & 0 \leq P^{\text{SUB}}_{0,t,q} \leq \sum_f P^{\text{SUB}}_{f,max}, \\
    &  0 \leq P^{\text{SUB}}_{1,t,q} \leq \sum_f P^{\text{SUB}}_{f,max}\Tilde{s}^{\text{feed}}_{f,t},\\
% %BESS
%     \sum_t P^{B}_{0,ch,t,q}-T^{ch}_{max}X^{B} \leq 0\\
%     SoC^{B/T}_{min}X^{B/T} \leq E^{B/T}_{0,t,q} \leq SoC^{B/T}_{max}X^{B/T}\\
%     0 \leq E^{B/T}_{1,t,q} \leq X^{B/T}\\
% %discharge and charge power limits
%     0 \leq P^{B/T}_{0/1,ch,t,q} \leq MZ^{B/T}_{0/1,ch,t,q}\\
%     0 \leq P^{B/T}_{0/1,dis,t,q} \leq M(1-Z^{B/T}_{0/1,ch,t,q})\\
%     P^{B/T}_{0/1,ch/dis,t,q} \leq coe^{B/T}_{ch/dis}X^{B/T}\\
% %Energy of ESS
%     E^{B/T}_{0/1,t,q} = E^{B/T}_{0/1,t-1,q}+P^{B/T}_{0/1,ch,t-1,q}\eta^{B/T}_{ch}\notag \qquad\qquad\qquad
%     \\\qquad\qquad\qquad\qquad\qquad\qquad -P^{B/T}_{0/1,dis,t-1,q}/\eta^{B/T}_{dis} \\
%% ESS-related cons
&Cons^{\text{ESS}}_{0,q}(X^{\text{B/T}}_E,X^{\text{B/T}}_P,P^{\text{B/T}}_{0,ch/dis,t,q},Z^{\text{B/T}}_{0,ch,t,q},E^{\text{B/T}}_{0,t,q}),\\
& Cons^{\text{ESS}}_{1,q}(X^{\text{B/T}}_E,X^{\text{B/T}}_P,P^{\text{B/T}}_{1,ch/dis,t,q},Z^{\text{B/T}}_{1,ch,t,q},E^{\text{B/T}}_{1,t,q}),\\
   & 0 \leq P^{\text{LS},d}_{1,t,q}, \\
    &X \in{\mathbb{X}}, Z^{\text{B/T}}_{0/1,ch,t,q} \in \{0,1\},\label{<Cons_MP_end>} \\
     &   \forall 1 \leq q \leq Q. \notag
\end{align}
\end{subequations}

The nature of the decomposition solving algorithm is to solve the master problem and subproblem iteratively until they converge. Here we use subscript $q$ to indicate the number of iterations. After each iteration, a new set of variables and constraints, which are also called \textit{cuts}, that are generated from the subproblem will be added back to the master problem. $\widetilde{\textbf{L}}$ represents the matrix of $\Tilde{l}^{d}_t$. It should be noticed that both $\Tilde{l}^{d}_t$ and $\Tilde{s}^{\text{Equi}}$ are constants instead of variables in the master problem, their value can be obtained from the solution of the subproblem. The subproblem can be shown as follows:
\begin{align}
    &\textbf{SP:   } \mathcal{Q}(\hat{X})=  \max_{l\in{\mathbb{L}},s\in{\mathbb{S}}}
    \min_{(Z,P)\in{F(x,l,s)}} f^{\text{ope}}_0+f^{\text{shed}}_1\\
    %m\sum_{t}[P^{SUB}_{t}r^{e}_{t}+(P^{CCHP}_{t}+P^{GB}_{t})r^{g}_{t}]
    &s.t.\quad \eqref{eq:con-balance-e-0}-\eqref{eq:con-Energy},(\ref{eq:con-balance-1})-(\ref{eq:con-ESS-1}),Z^{\text{B/T}}_{0/1,ch,t} \in \{0,1\}, l\in{\mathbb{L}},s\in{\mathbb{S}} \notag,
\end{align}
where $\hat{X}$ is the current optimal planning decision from the master problem. Nested C\&CG algorithm contains an outer layer algorithm and an inner layer algorithm to solve the master problem and the subproblem respectively. 
\subsection{Outer Layer Algorithm}
On the basis that the subproblem can be solved by the inner layer algorithm, the outer layer algorithm can be applied to solve the proposed planning model in the following steps:
\begin{itemize}
  \item[1)]\textit{Step 1:} Set $LB=-\infty$, $UB=+\infty$, $\epsilon = 0.0001$ and $Q=0$.  
  
  \item[2)]\textit{Step 2:} Solve the master problem \textbf{MP}.

Derive an optimal solution $(X^{*},\psi^{*})$ and update $LB=\sum_{\text{Equi}}\sum_{n}IC^{\text{Equi}}_{n} X^{\text{Equi},*}_n+\sum_{\text{ESS}}IC^{\text{ESS}}_E X^{\text{ESS},*}_E+IC^{\text{ESS}}_P X^{\text{ESS},*}_P+\psi^* $. If $UB-LB \leq \epsilon$, return the solution and terminate. 

  \item[3)]\textit{Step 3:} Obtain $\hat{X}$ from step 2 and solve the subproblem \textbf{SP} by inner layer algorithm. Derive an optimal solution $(l^{*},s^{*})$ and update $UB=min\{UB,\sum_{\text{Equi}}\sum_{n}IC^{\text{Equi}}_{n} X^{\text{Equi},*}_n+\sum_{\text{ESS}}IC^{\text{ESS}}_E X^{\text{ESS},*}_E +IC^{\text{ESS}}_P X^{\text{ESS},*}_P+\mathcal{Q^*}(\hat{X})\}$. If $UB-LB \leq \epsilon$, return and terminate. The details of the solution process for the subproblem can be found in inner level algorithm.

  \item[4)]\textit{Step 4:}  Obtain $(\Tilde{l}$,$\Tilde{s})$ from subproblem, update $Q=Q+1$. Create variables $(P^{q+1}_{0/1},Z^{q+1}_{0/1})$ and add constraints \eqref{<Cons_MP_begin>}$-$\eqref{<Cons_MP_end>} with index $q=q+1$ to the \textbf{MP}, go to \textit{Step 2}.
\end{itemize}

% \textcolor{blue}{
%  The pseudocode of outer layer algorithm can be found in Algorithm \ref{algo:OLA}. }
% \renewcommand{\algorithmicrequire}{\textbf{Input: }\unskip}
% \renewcommand{\algorithmicensure}{\textbf{Output: }\unskip}

% \textcolor{blue}{
% \begin{algorithm}[H]
%   \caption{Outer layer C\&CG}
%   \label{algo:OLA}
%   \small
%   \begin{algorithmic}[1]
%     \color{blue} 
%     \REQUIRE $LB = -\inf, UB = +\inf, \epsilon = 0.0001, Q=0$
%     \ENSURE $X^*,Z^*,P^*$
%     \WHILE{$UB-LB > \epsilon$}
%     \STATE Solve the MP with ($\tilde{l},\tilde{s}$)
%     \STATE Update $LB \rightarrow \displaystyle \sum_{\text{Equi}} \sum_{n}IC^{\text{Equi}}_{n} X^{\text{Equi},*}_n+\sum_{\text{ESS}}IC^{\text{ESS}}X^{\text{ESS},*}+\psi^*$
%     \STATE Solve the \textbf{SP} with $\hat{X}$ by Algorithm \ref{algo:ILA}
%     \STATE Update $UB \rightarrow min\{UB, \displaystyle \sum_{\text{Equi}}\sum_{n}IC^{\text{Equi}}_{n} X^{\text{Equi},*}_n+\sum_{\text{ESS}}IC^{\text{ESS}}X^{\text{ESS},*}+ \mathcal{Q}^*(\hat{X})\}$
%     \STATE Create variables ($P^{q+1}_{0/1},Z^{q+1}_{0/1}$) and add \eqref{<Cons_MP_begin>}$-$\eqref{<Cons_MP_end>} back to \textbf{MP}
%     \STATE Update $Q \rightarrow Q+1$
%     \ENDWHILE
%   \end{algorithmic}
% \end{algorithm}
% }

\subsection{Inner Layer Algorithm}
As shown in \textit{Step 3} of the outer layer algorithm, the \textbf{SP} is solved by the inner layer algorithm. To better illustrate the inner layer algorithm, we reformulate the subproblem and get its tri-level equivalent form as follows:
\begin{equation}
    \mathcal{Q}(\hat{X}) = \max_{l\in{\mathbb{L}},s\in{\mathbb{S}}} \min_{Z \in \Phi_{Z}}
    \min_{P} f^{\text{ope}}_0+f^{\text{shed}}_1 \label{eq:subproblem}
   %\min_{P \in \Phi_{P}(\hat{X},l,s,Z)} f^{ope}+f^{shed}
\end{equation}
where $\Phi_{Z}=\{Z^{r}\}^{R}_{r=1}$. By making use of the countability of $\Phi_{Z}$, we have
\begin{gather}
    \mathcal{Q}(\hat{X}) = \max \sigma,\\
     s.t. \quad \sigma \leq 
     \min\{ m\sum_{t}[P^{\text{SUB}}_{0,t,r}\mathcal{R}^{e}_{t}+(\sum_i P^{\text{CCHP}}_{0,i,t,r}+\sum_jP^{\text{GB}}_{0,j,t,r}\notag \\)\mathcal{R}^{g}_{t}
     +P^{\text{LS},d}_{1,t,r}p^{\text{LS},d}_{t}]
    : Cons^{\text{SP}}_{r},\quad r=1,...,R\}. \label{<theta>}
\end{gather}

Here $r$ is used to denote the number of inner algorithm iterations. The constraints $Cons^{\text{SP}}_r$ are almost the same as the constraints of the original optimization problem except with $\hat{X}$ as constant. In standard nested C\&CG method, the minimization problem on the right side of \eqref{<theta>} will be converted into a feasible problem by KKT conditions directly. In contrast, we employ the strong duality theory to reformulate this minimization problem into a maximization problem, which is particularly effective for solving our planning model. This formulation will be shown in sequel. The formulation based on KKT conditions is also provided in the Appendix.

\textcolor{blue}{} This subproblem will also be reformulated into a master problem and a subproblem and solved by a decomposition algorithm.  The inner layer algorithm follows steps as:

\begin{itemize}
  \item[1)]\textit{Step 1:}  Set $LB_s=-\infty$, $UB_s=+\infty$, $\epsilon_s = 0.0001$ and $R$ = 0.
  
\item[2)]\textit{Step 2:} Solve the master problem of subproblem:
  \begin{subequations}
  \begin{align}
  &\textbf{MPs:   } \max \sigma, \\
&s.t. \quad \sigma \leq -T^{ch}_{max}\hat{X}^{\text{B}}_E\pi^6_r+\sum_{t}\bigg(l^e_t \pi^1_{t,r}+l^h_t \pi^2_{t,r}+l^c_t \pi^3_{t,r}- \notag \\
&\sum_n P^{\text{Equi}}_{n,max}\hat{X}^{\text{Equi}}_{n}\pi^4_{n,t,r}-\sum_f P^{\text{SUB}}_{f,max}\pi^5_{t,r}+\text{SoC}^{\text{B}}_{min}\hat{X}^{\text{B}}_E\pi^7_{t,r}\notag \\
&-\text{SoC}^{\text{B}}_{max}\hat{X}^{\text{B}}_E\pi^8_{t,r} +\text{SoC}^{\text{T}}_{min}\hat{X}^{\text{T}}_E\pi^9_{t,r}
-\text{SoC}^{\text{T}}_{max}\hat{X}^{\text{T}}_E\pi^{10}_{t,r}- \notag \\
&\hat{X}^{\text{B}}_P\tilde{Z}^{\text{B}}_{0,ch,t,r}\pi^{11}_{t,r} -\hat{X}^{\text{T}}_P\tilde{Z}^{\text{T}}_{0,ch,t,r}\pi^{12}_{t,r}-\hat{X}^{\text{B}}_P(1-\tilde{Z}^{\text{B}}_{0,ch,t,r})\pi^{13}_{t,r} \notag \\
& -\hat{X}^{\text{T}}_P(1-\tilde{Z}^{\text{T}}_{0,ch,t,r})\pi^{14}_{t,r}+l^e_t \pi^{17}_{t,r}+l^h_t \pi^{18}_{t,r}+l^c_t \pi^{19}_{t,r} \notag \\
& -\sum_n P^{\text{Equi}}_{n,max}s^{\text{Equi}}_{n,t}\hat{X}^{\text{Equi}}_{n}\pi^{20}_{n,t,r}-\sum_f P^{\text{SUB}}_{f,max}s^{\text{Feed}}_{f,t}\pi^{21}_{t,r}\notag \\ 
&-\hat{X}^{\text{B}}_E\pi^{22}_{t,r}-\hat{X}^{\text{T}}_E\pi^{23}_{t,r}-\hat{X}^{\text{B}}_P\tilde{Z}^{\text{B}}_{1,ch,t,r}\pi^{24}_{t,r}-\hat{X}^{\text{T}}_P\tilde{Z}^{\text{T}}_{1,ch,t,r}\pi^{25}_{t,r}  \notag \\
& -\hat{X}^{\text{B}}_P(1-\tilde{Z}^{\text{B}}_{1,ch,t,r})\pi^{26}_{t,r}-\hat{X}^{\text{T}}_P(1-\tilde{Z}^{\text{T}}_{1,ch,t,r})\pi^{27}_{t,r}\bigg), \label{<Cons_MPs_begin>}\\
&\pi^1_{t,r},\ldots,\pi^{14}_{t,r},\pi^{17}_{t,r},\ldots,\pi^{27}_{t,r}  \geq 0,\quad \pi^{15,16,28,29}_{t,r} \in \mathbb{R} \label{eq:MPs-dual}\\
    &Cons^{SD}_r,\label{eq:MPs-SD}\\  
     &l \in{\mathbb{L}}, s \in{\mathbb{S}},\\
    &\forall 1 \leq r \leq R .\notag
  \end{align}
\end{subequations}
The inner layer algorithm is also iterative, new variables and constraints with subscript $r$ are added to the \textbf{MPs} with each iteration. It should be noticed that $\hat{X}$ and $\tilde{Z}$ are constants rather than variables in \textbf{MPs}. The former one comes from the solution of \textbf{MP}, and the latter one comes from the solution of subproblem of \textbf{SP}. The right-hand side of \eqref{<Cons_MPs_begin>} represents the objective function of the dual problem obtained by reformulating \eqref{<theta>} using the SD theory. Since the reformulated maximization problem lies on the right-hand side of the inequality, the "max" can be omitted directly. Now the decision variables become dual variable $\pi$, each corresponds to a constraint in primal problem, i.e., $Cons^{\text{SP}}_{r}$ in \textbf{SP}. Constraint (\ref{eq:MPs-dual}) is for these dual variables. $Cons^{\text{SD}}_{r}$ represents other constraints of the dual problem. They can be simply derived from the primal problem so we will not show them here.

Unlike the tasks typically handled by nested C\&CG, both uncertainty sets $\mathbb{L}$ and $\mathbb{S}$ in our IES planning model contain binary variables. This will introduce terms like $l^e_t \pi^1_{t,r}$ and $P^{\text{Equi}}_{n,max}s^{\text{Equi}}_{n,t}\hat{X}^{\text{Equi}}_{n}\pi^{20}_{n,t,r}$ to \textbf{MPs} and significantly increases the non-convexity. Here, we propose an exact linearization method to enhance solution efficiency. Take $l^e_t \pi^1_{t,r}$ as example:
\begin{align*}
    l^{e}_t \pi^1_{t,r} &= (\bar{l}^{e}_t+\Delta l^{e}_t u^e_{t,+}  - \Delta l^{e}_tu^e_{t,-})\pi^1_{t,r} \\
    &=\bar{l}^{e}_t \pi^1_{t,r}+\Delta l^{e}_t\omega^+_{t,r}-\Delta l^{e}_t\omega^-_{t,r},
\end{align*}
where $\omega^+_{t,r}:=\pi^1_{t,r}u^e_{t,+}$ satisfies
\begin{subequations}
\begin{gather}
    \omega^+_{t,r} \leq M \cdot u^e_{t,+},\\
    \omega^+_{t,r} \leq \pi^1_{t,r},\\
    \omega^+_{t,r} \geq \pi^1_{t,r}-M(1-u^e_{t,+}),\label{reform}\\
    \omega^+_{t,r} \geq 0. 
\end{gather}\label{general-reform}
\end{subequations}
Here $M$ is a big constant. For $\omega^-_{t,r}=\pi^1_{t,r}u^e_{t,-}$, it also has similar constraints. It is worth noting that binary variables $u^e_{t,+}$ and $u^e_{t,-}$ from load uncertainty set are subject to a mutual exclusivity constraint, i.e., $u^e_{t,+}+u^e_{t,-} \leq 1$. Therefore, we can combine the \eqref{reform} and the corresponding constraint for $u^e_{t,-}$ into the following equivalent expression:
\begin{equation}
    \omega^+_{t,r}+\omega^-_{t,r} \geq \pi^1_{t,r}-M(1-u^e_{t,+}-u^e_{t,-}).\label{new-reform}
\end{equation}
This further reformulation method will significantly enhance computational efficiency. The same method as \eqref{general-reform} can be used to recast other terms containing the multiplication of continuous and binary variables. After the reformulation, \textbf{MPs} becomes an MILP problem rather than a mixed-integer quadratically constrained programming (MICQP) problem, which is typically the case in standard nested C\&CG method and is more intractable.

After solving the \textbf{MPs}, derive an optimal solution $(l^*,s^*)$ and update $UB_s=\sigma^*$. If $UB_s-LB_s \leq \epsilon_s$, return and terminate.
  \item[3)]\textit{Step 3:} Obtain $(\hat{l},\hat{s})$ from \textit{Step 2} and solve the subproblem of \textbf{SP}:
  \begin{subequations}
  \begin{align}
      &\textbf{SPs:}
  \min_{(Z,P)\in{F(\hat{x},\hat{l},\hat{s})}} m\sum_{t}[P^{\text{SUB}}_{0,t,r}\mathcal{R}^{e}_{t}+(\sum_i P^{\text{CCHP}}_{0,i,t,r}+\notag \\
  & \sum_j P^{\text{GB}}_{0,j,t,r})\mathcal{R}^{g}_{t}
    +P^{\text{LS},d}_{1,t,r}p^{\text{LS},d}_{t}],\\
   &s.t.\quad\widehat{\textbf{L}} \leq \textbf{C}^{\text{SUB}}\textbf{P}^{\text{SUB}}_{0,r}+\sum_{\text{Equi}}\textbf{C}^{\text{Equi}}\textbf{P}^{\text{Equi}}_{0,r}-\textbf{P}^{\text{ESS}}_{0,r}, \\
   & \widehat{\textbf{L}} \leq \textbf{C}^{\text{SUB}}\textbf{P}^{\text{SUB}}_{1,r}+\sum_{\text{Equi}}\textbf{C}^{\text{Equi}}\textbf{P}^{\text{Equi}}_{1,r}-\textbf{P}^{\text{ESS}}_{1,r}+\textbf{P}^{\text{LS}}_{1,r},\\
&0 \leq P^{\text{Equi}}_{0,n,t,r}
    \leq P^{\text{Equi}}_{n,max}\hat{X}^{\text{Equi}}_{n},\\
&0 \leq P^{\text{Equi}}_{1,n,t,r} \leq P^{\text{Equi}}_{n,max}\hat{s}^{\text{Equi}}_{n,t}\hat{X}^{\text{Equi}}_{n},\\
 & 0 \leq P^{\text{SUB}}_{0,t,r} \leq \sum_f P^{\text{SUB}}_{f,max}, \\
    &  0 \leq P^{\text{SUB}}_{1,t,r} \leq \sum_f P^{\text{SUB}}_{f,max}\hat{s}^{\text{feed}}_{f,t},\\
&Cons^{\text{ESS}}_{0,r}(\hat{X}^{\text{B/T}}_E,\hat{X}^{\text{B/T}}_P,P^{\text{B/T}}_{0,ch/dis,t,r},Z^{\text{B/T}}_{0,ch,t,r},E^{\text{B/T}}_{0,t,r}),\\
& Cons^{\text{ESS}}_{1,r}(\hat{X}^{\text{B/T}},\hat{X}^{\text{B/T}}_P,P^{\text{B/T}}_{1,ch/dis,t,r},Z^{\text{B/T}}_{1,ch,t,r},E^{\text{B/T}}_{1,t,r}),\label{<Cons_SPs_Cons_ESS>}\\
&0 \leq P^{\text{LS},d}_{1,t,r}, \label{<Cons_SPs_Cons_LS>}\\
&Z^{\text{B/T}}_{0/1,ch,t,q} \in \{0,1\}.
  \end{align}
  \end{subequations}

The \textbf{SPs} tends to optimize the second-stage variable $Z$ based on given $ \widehat{\textbf{L}} $, $\hat{s}^{\text{Equi}}$ and $\hat{s}^{\text{Feed}}$ from \textbf{MPs}, and $\hat{X}^{\text{Equi}}$, $\hat{X}^{\text{B/T}}_{E/P}$ from \textbf{MP}.
 
 After solving \textbf{SPs}, derive an optimal solution $(Z^*,P^*)$ and update $LB_s=\max\{LB_s,m\sum_{t}[P^{\text{SUB},*}_{0,t,r}\mathcal{R}^{e}_{t}+(\sum_i P^{\text{CCHP},*}_{0,i,t,r}+\sum_j P^{\text{GB},*}_{0,j,t,r})\mathcal{R}^{g}_{t}]+P^{\text{LS},d,*}_{1,t,r}p^{\text{LS},d}_t\}$. If $UB_s-LB_s \leq \epsilon_s$, return and terminate.
  
  \item[4)]\textit{Step 4:}  
  Obtain $\Tilde{Z}$ from \textbf{SPs}, update $R=R+1$. Create variables $P^{r+1}$ and add constraints \eqref{<Cons_MPs_begin>}$-$\eqref{eq:MPs-SD} with index $r=r+1$ to the \textbf{MPs}, go to \textit{Step 2}.
\end{itemize}

\begin{figure}[H]
  \centering
  \includegraphics[width=1\linewidth]{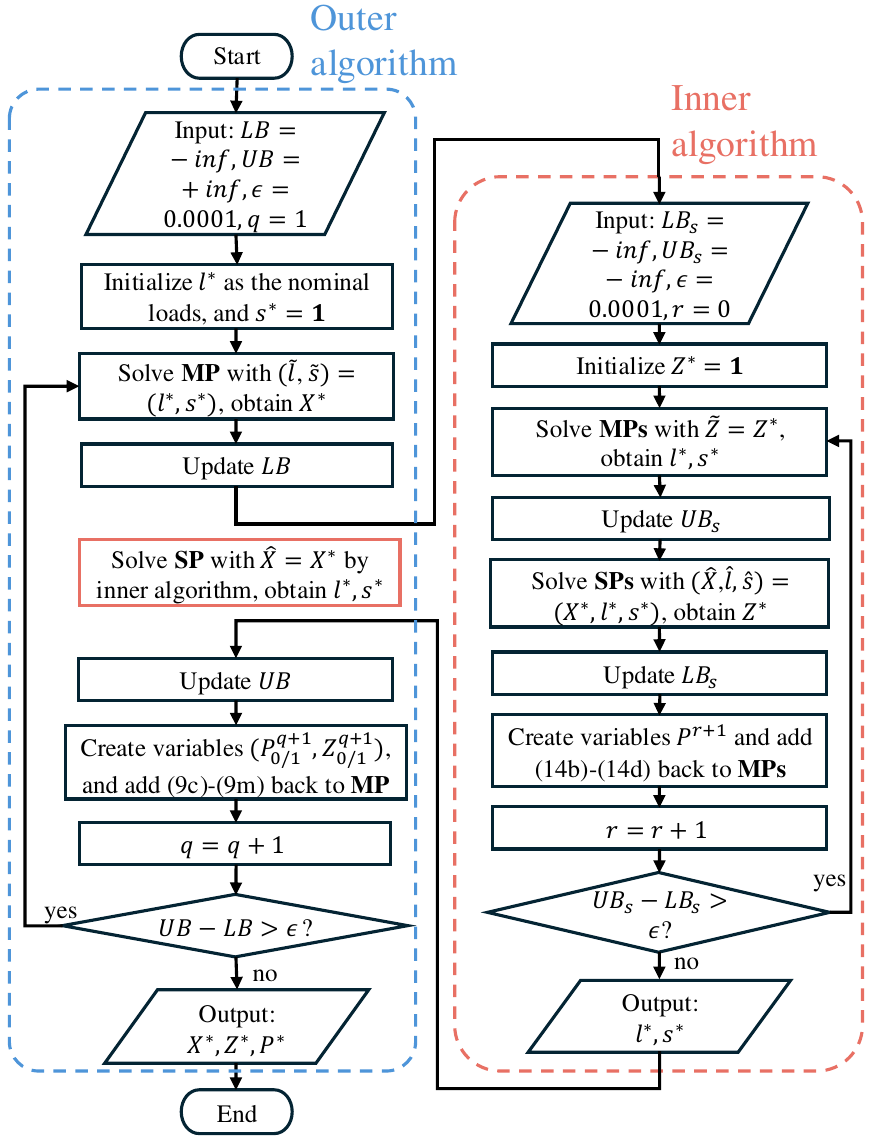}
  \caption{Flowchart of the solving algorithm.}
  \label{fig:flowchart}
\end{figure}

The flowchart of the whole solving algorithm is shown in Fig. \ref{fig:flowchart}.

% \textcolor{blue}{
% Please refer to Algorithm \ref{algo:ILA} for a more concise solution process. 
% \begin{algorithm}[H]
%   \caption{Inner layer C\&CG}
%   \label{algo:ILA}
%   \small
%   \begin{algorithmic}[1]
%   \color{blue}
%     \REQUIRE $LB_s = -\inf, UB_s = +\inf, \epsilon_s = 0.0001, R=0$
%     \ENSURE $l^*,s^*$
%     \WHILE{$UB_s-LB_s \leq \epsilon_s$}
%     \STATE Solve the \textbf{MPs} with $\tilde{Z}$
%     \STATE Update $UB_s \rightarrow \theta^*$
%     \STATE Solve the \textbf{SPs} with ($\hat{X},\hat{l},\hat{s}$) 
%     \STATE Update $LB_s \rightarrow max\{LB_s, \sigma^*\}$
%     \STATE Create variables $P^{r+1}$ and add constraints back to \textbf{MPs}
%     \STATE Update $R \rightarrow R+1$
%     \ENDWHILE
%   \end{algorithmic}
% \end{algorithm}}

\section{Case Study}
% In order to verify the effectiveness of the proposed planning model, an industrial park is selected as the numerical case. In this section, we first analyze the planning results obtained from the proposed planning model and observe its variation with different uncertainty budgets. Then, the Monte Carlo simulation (MCS) method is used to evaluate the reliability of the system. The role of ESS in improving reliability is also analyzed. At last, the convergence performance of the nested C\&CG algorithm based on KKT conditions and strong duality theory will be analyzed and compared. 

\subsection{Case Condition}
\begin{figure}
  \centering
  \includegraphics[width=1\linewidth]{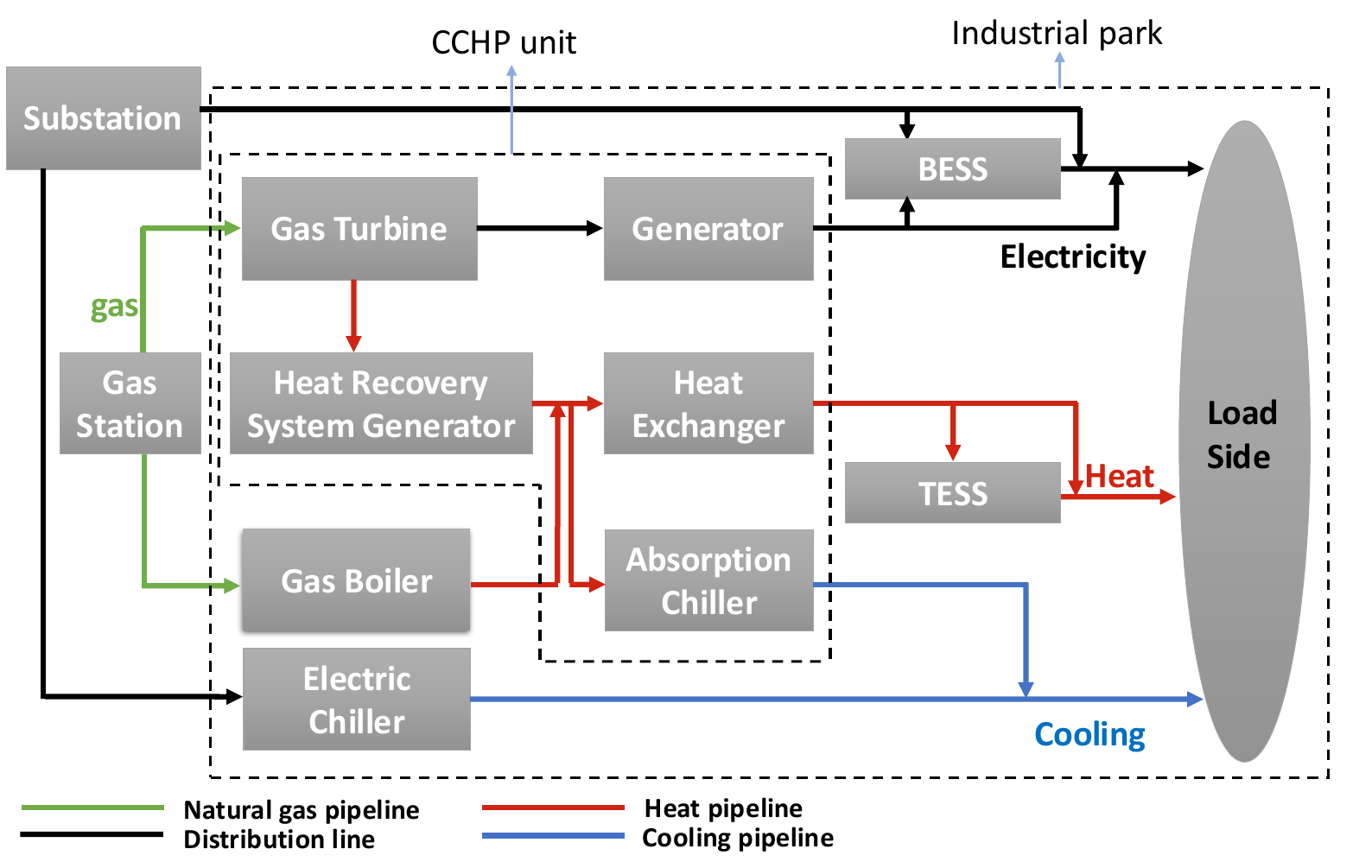}
  \caption{Schematic of an industrial park.}
  \label{fig:park}
\end{figure}

A real industrial park is used as the numerical case. Substation and natural gas station outside the park provide electricity and natural gas respectively as input for the park. Electricity, heat, and cooling energy are needed on the demand side. Four typical days with various load demand data are generated by K-means as a baseline \cite{RN294}. The maximum electric, heat, and cooling loads of the park are 60MW, 85MW, and 75MW respectively. The equipment to be selected in the park includes CCHP units, gas boilers, and electric chillers, of which there are 7 types of CCHP units, 10 types of gas boilers and electric chillers. The substation supplies power through 5 distribution lines, each of 10MVA capacity. The parameters of equipment can be found in relevant study \cite{RN97}. For ESS, we consider both energy capacity and power capacity of BESS. Costs of lithium-ion batteries can be found in recent report \cite{RNweb}, and study \cite{RN327}. In BESS, cost for both energy and power capacity must be considered, as the inverter, which determines the power capacity, is expensive. However, for TESS, the primary cost is associated with energy capacity. Therefore, we only consider its energy capacity and set half of its capacity as the power limit.
The specific equipment connection of the park is shown in Fig. \ref{fig:park}. 

To formulate proper component contingency uncertainty set for the proposed two-stage robust planning model, we refer to relevant parameters in other studies \cite{RN203}, i.e. the failure and repair rates of different component. We set $\Gamma^{N}$ as 1, which means we mainly focus on the operation scenarios with at most one contingency in the same period. Otherwise the results will be too conservative. For other parameters, although there is no standard approach to convert these annualized reliability parameters to the parameters of typical days, we adjust these uncertainty budgets to assess their impact on planning results and discuss the rational of parameter selection in next section.

\subsection{Analysis on Planning Results}
After the experiment, we observed that the duration of equipment contingency, $\Gamma^{D}_{n}$, significantly affects the final planning scheme, whereas the interval between two equipment contingencies, $\Gamma^{I}_{n}$, has no impact on the result. So, here we set $\Gamma^{I}_{n}$ as $0.5*\text{Period}$ for all following cases. We will further discuss this in next section. We observe planning schemes variations for different $\Gamma^{D}_{n}$ values in the following cases:
\begin{itemize}
  \item[1)]\textit{Case 1:} Deterministic planning model, i.e., $\Gamma^{D}_{n}=0$;

  \item[2)]\textit{Case 2:} Two-stage robust planning model with uncertainty budget: \{$\Gamma^{D}_{n} = 3$\};

  \item[3)]\textit{Case 3:} Two-stage robust planning model with uncertainty budget: \{$\Gamma^{D}_{n} \geq 6$\};
  
  We also consider uncertain loads in the following case: 
  \item[4)]\textit{Case 4:} Two-stage robust planning model with uncertainty budget: \{$\Gamma^{D}_{n} \geq 6$, $\Gamma^{l}=\text{Period}$, $\Delta l^{d}_t=2\%*\bar{l}^{d}_t $\};
\end{itemize}

The planning results are shown in Table \ref{tab:planning result CCHP}-\ref{tab:planning result ESS}. In these tables, ‘1’ indicates equipment construction, while ‘0’ indicates no construction. 
\begin{table}
  \centering
  \caption{The planning results for CCHP units. }
  \begin{tabular}{ccccccccc}
    \hline
    \hline
    Capacity   & 5 & 10 & 15 & 20 & 25 & 30 & 35 & Total/MW\\
    \hline
    \textit{Case 1,2}   & 0 & 0 & 1 & 0 & 0 & 0 & 1 & 50\\
    \textit{Case 3,4}   & 0 & 0 & 0 & 1 & 0 & 1 & 0 & 50\\
    \hline
    \hline
  \end{tabular}
  \label{tab:planning result CCHP}
\end{table}
As for CCHP units, all cases invest 50MW in total. However, the first two cases choose two CCHP units with 15MW and 35MW respectively while the last two csaes choose those with 20MW and 30MW. This is because when any CCHP unit fails, the remaining equipment should be able to supply sufficient load demand. As a result, the planning methods considering contingencies with longer period tend to make the capacity of two units more balanced.
\begin{table}
  \centering
  \caption{The planning results for gas boilers. }
  \begin{tabular}{cccccccccc}
    \hline
    \hline
    Capacity   &  10 & 20 & 30 & 40 & 50 & 60 & 70 & 80-100 & Total/MW\\
    \hline
    \textit{Case 1}   & 0& 0& 0& 0& 1& 0& 0& 0& 50\\
    \textit{Case 2}   & 1& 0& 1& 0& 0& 0& 0& 0& 40\\
    \textit{Case 3,4}   & 1& 1& 1& 0& 0& 0& 0& 0& 60\\
    \hline
    \hline
  \end{tabular}
  \label{tab:planning result GB}
\end{table}

Regarding gas boilers, only one single gas boiler of 50 MW capacity is constructed under the deterministic planning scheme. In contrast, the robust planning schemes replace the large capacity gas boiler with several gas boilers in relatively small capacity to mitigate potential shedding load caused by single equipment contingency. This necessitates an expansion of the total capacity of GB or TESS to improve IES reliability. To be more specific, when relatively short failure is considered, \textit{Case 2} chooses to replace the 50MW boiler with two boilers of 10MW and 30MW. The total capacity of GB decreases and larger TESS is invested. The role of TESS will be analyzed in the next part. When considering contingency for more than 3 hours, merely replacing the larger boiler with smaller ones is insufficient. So, \textit{Case 3 and 4} both choose to invest in three boilers of 10MW, 20MW, and 30MW in order to avoid the huge potential load shedding caused by the failure of each single unit. They also choose to enlarge the capacity of TESS compared with \textit{Case 1}.

\begin{table*}
  \centering
  \caption{The planning results for electric chillers. }
  \begin{tabular}{cccccccccccc}
    \hline
    \hline
    Capacity   &  2.5 & 5.0 & 7.5 & 10.0 & 12.5 & 15.0 & 17.5 & 20.0 & 22.5 & 25.0 & Total/MW\\
    \hline
    \textit{Case 1}    & 0 & 0 & 0 & 0 & 0 & 0 & 0 & 0 & 1 & 0 & 22.5\\
    \textit{Case 2}   & 0 & 0 & 0 & 0 & 1 & 0 & 0 & 0 & 1 & 1 & 60\\
    \textit{Case 3,4}  & 0 & 0 & 1 & 0 & 0 & 0 & 0 & 0 & 1 & 1 & 55\\
    \hline
    \hline
  \end{tabular}
  \label{tab:planning result EC}
\end{table*}

The planning schemes of electric chillers under deterministic model is too aggressive. So when uncertain equipment contingency is considered, the total capacity of EC should be increased to ensure the normal cooling energy supply. Additionally, since the CCHP units selected under \textit{Case 3} and \textit{Case 4} are more stable, the total capacity of EC installed in these cases has decreased compared to \textit{Case 2}.
\begin{table}
  \centering
  \caption{The planning results for ESS (MWh/MW)}.
  \begin{tabular}{ccccc}
    \hline
    \hline
       & \textit{Case 1} & \textit{Case 2} & \textit{Case 3} & \textit{Case 4}\\
    \hline
    BESS   & 7.20/2.13 & 27.03/8.62 & 7.80/2.28 & 14.24/4.22\\
    TESS   & 3.08 & 89.24 & 35.78 & 44.83\\
    \hline
    \hline
  \end{tabular}
  \label{tab:planning result ESS}
\end{table}
As for the planning schemes of ESS, even the contingency is not considered, \textit{Case 1} chooses to invest 7.2 MWh BESS and 3.08MWh TESS to earn profit from the time-of-use energy tariff. In other cases, the capacity of both invested BESS and TESS are significantly affected by the uncertainty budget $\Gamma^D_n$, i.e., the duration of equipment contingency. We will analyze it later.

The total cost of different planning schemes can be shown in Table \ref{tab:cost}. Obviously, the planning schemes from robust models increase the investment cost to improve reliability. It should also be noted that even the most robust planning scheme, \textit{Case 4}, only increases the total cost by around $2\%$ compared to the deterministic scheme. To better compare these planning schemes, we will evaluate their reliability in the next part.
\begin{table}
\centering
\caption{Cost of different planning schemes (unit: ¥$10^4$). }
\begin{tabular}{ccccllcc}
\hline
\hline
\multirow{2}{*}{} & \multicolumn{2}{c}{Investment}                                                & \multicolumn{3}{c}{\multirow{2}{*}{\begin{tabular}[c]{@{}c@{}}Operational \\ cost\end{tabular}}} & \multicolumn{2}{c}{Total}  \\                                      \cline{2-3} \cline{7-8} 
                & Cost    & Increment & \multicolumn{3}{c}{} & Cost     & Increment \\\hline
\textit{Case 1} & 58312.6 & -      & \multicolumn{3}{c}{183655.5} & 241968.1 & - \\
\textit{Case 2} & 66396.5 & 13.86\% & \multicolumn{3}{c}{179239.7} & 245636.2 & 1.52\% \\
\textit{Case 3} & 61516.9 & 5.50\% & \multicolumn{3}{c}{184642.2} & 246159.1 & 1.73\% \\
\textit{Case 4} & 62974.7 & 7.99\% & \multicolumn{3}{c}{184449.5} & 247424.2 & 2.25\% \\
\hline
\hline
\end{tabular}
\label{tab:cost}
\end{table}

\subsection{Reliability Assessment}
The MCS method which is simple and efficient is applied to quantify the reliability level of the planning schemes obtained above. The reliability indices we use include EENS, LOLF, and LOLE, which can well reflect the total amount, frequency and duration of different types of energy load shedding. MCS involves simulating numerous operational scenarios to reflect the actual system performance, with each scenario incorporating variations in electricity, heat, and cooling demands, which follow a normal distribution with fluctuations of $\pm$5\%, $\pm$8\%, and $\pm$5\% respectively. 
Additionally, to simulate the operating states of each invested equipment, we use a two-status model, where equipment transitions between ``normal" and ``fault" states, with transformation probabilities governed by parameters such as equipment failure rates and repair rates. Detailed methods and relevant parameters can be found in \cite{RN179, RN203}, \cite{RN344}. 
% In this way, a large number of IES operation scenarios are generated, based on which we evaluate the reliability of the obtained planning schemes.
We choose the sample size as 500 years (each year with 8760 hours) to calculate these reliability indices. The evaluation results of obtained planning schemes are shown in Table \ref{tab:reliability result}. 
\begin{table}
\centering
\caption{The results of reliability evaluation.}
\begin{tabular}{ccccc}
\hline
\hline
                         & \textit{Case} & \begin{tabular}[c]{@{}c@{}}EENS\\ (MW/year)\end{tabular} & \begin{tabular}[c]{@{}c@{}}LOLE\\ (hours/year)\end{tabular} & \begin{tabular}[c]{@{}c@{}}LOLF\\ (occ/year)\end{tabular} \\ \hline
\multirow{3}{*}{Electricity}    & \textit{1}    & 0.07 & 0.02 & 0.01 \\
                         & \textit{2}    &  0.02 & 0.00 & 0.00 \\
                         & \textit{3,4}    &  0.00 & 0.00 & 0.00 \\ \hline
\multirow{4}{*}{Heat}    & \textit{1}    & 70.49 & 6.58 & 3.21 \\
                         & \textit{2}    &  3.74 & 0.31 & 0.10 \\
                         & \textit{3}    &  0.15 & 0.03 & 0.01 \\
                         & \textit{4}    &  0.07 & 0.01 & 0.01 \\ \hline
\multirow{4}{*}{Cooling} & \textit{1}    & 25.24 & 3.88 & 2.37 \\
                         & \textit{2}    &  0.12 & 0.02 & 0.01 \\ 
                          & \textit{3}    &  0.06 & 0.01 & 0.01 \\ 
                           & \textit{4}    &  0.00 & 0.00 & 0.00 \\ 
                         \hline\hline
\end{tabular}
\label{tab:reliability result}
\end{table}

Compared to other two types of loads, the electricity supply is relatively reliable as the substation outside the IES can supply power. However, there will be severe thermal load shedding as well as cooling load shedding in \textit{Case 1}. The frequency of load shedding occurrence within a year is also high. This is unacceptable for most IES, especially for IES with high-reliability requirements, where a small amount of load shedding can result in significant economic losses. The reason for this result is that this case does not consider possible equipment contingencies and load fluctuations. Therefore leads to an unreliable planning scheme. 

By considering potential uncertainties, \textit{Case 2-4} significantly reduce all three relevant indices of both heat and cooling load. Compared to \textit{Case 1}, \textit{Case 2} enlarges the capacity of TESS and reduces the heat EENS, LOLE, and LOLF by $94.7\%$, $95.3\%$ and $96.9\%$ respectively. \textit{Case 3 and 4} reduce all kinds of shedding load to almost zero. From the last part, these two cases improve the system reliability by only increasing around $1\%-2\%$ total cost. 

\subsection{Analysis on ESS}
Many studies have extensively examined the economic role of ESS, such as their participation in direct arbitrage within energy markets and provision of ancillary services \cite{RN312}. However, research on their impact in improving reliability, particularly in terms of concrete quantification, remains limited. In this subsection we provide a detailed analysis based on our planning result for ESS.

The capacity of BESS and the selection of CCHP invested under different uncertainty budgets are given in Table \ref{tab:capacity of BESS}. The capacity of the CCHP determines the electricity generation capability of the IES, as the substation capacity is fixed. Comparing cases with the same CCHP selection, e.g., when $\Gamma^D_n=0,1,2,3$, we find that the longer the failure duration considered, the larger the investment in BESS capacity. This indicates that the BESS can be used to mitigate potential electricity load shedding. It also should be noticed that the energy power ratio of the BESS is roughly equal to the contingency duration we consider. For example, when $\Gamma^D_n=2$, we choose BESS with 17.95MWh and 8.57MW. This ensures that when a two-hour failure which can potentially cause electricity shortage occurs, the BESS can support the IES and maintain power supply. When we set $\Gamma^D_n \geq 4$, a more reliable scheme of CCHP investment is selected, and the BESS capacity returns to a level similar to that of the deterministic scenario. In this case, the main role of BESS becomes earning energy arbitrage again. Moreover, the last two cases have the same selections for all equipment except ESS. The case that consider load uncertainty invests nearly twice as much in BESS capacity compared to the case that does not. This highlights the important role of BESS in mitigating load fluctuations.

\begin{table}
  \centering
  \caption{The capacity of BESS and CCHP under different $\Gamma^D_n$.}
  \begin{tabular}{cccccc}
    \hline
    \hline
       $\Gamma^D_n$ & \begin{tabular}[c]{@{}c@{}}BESS\\ /MWh/MW\end{tabular} & \begin{tabular}[c]{@{}c@{}}Energy\\ power ratio\end{tabular} & \begin{tabular}[c]{@{}c@{}}CCHP\\ /MW \end{tabular}\\
    \hline
    0   & 7.20/2.13 & 3.38 & 15+35\\
    1 & 11.96/8.57 & 1.39 & 15+35\\
    2 & 17.95/8.57 & 2.09 & 15+35\\
    3   & 27.03/8.62 & 3.13 & 15+35\\
    4-6 & 7.80/2.28 & 3.42 & 20+30\\
    6 (with uncertain load) & 14.24/4.22 & 3.37 & 20+30\\
    \hline
    \hline
  \end{tabular}
  \label{tab:capacity of BESS}
\end{table}

Similarly, the planning results of TESS and GB under different uncertainty budgets are given in Table \ref{tab:capacity of TESS and GB}. In the deterministic model, i.e., $\Gamma^D_n=0$, 3.08MW TESS is invested. It can be used to save operational costs by storing heat surplus from CCHP and releasing it when heat is needed. This is the first function of TESS, and also the most commonly discussed one. When relatively short equipment contingency, which is not longer than 3 hours, is considered, large capacity TESS will be invested. On the one hand, large capacity TESS is used to provide reserve to against short-term failures of other equipment. On the other hand, it can fill the heat supply gap caused by the reduction in total GB capacity compared to the deterministic planning scheme. When longer periods of contingency are considered, raising the capacity of TESS alone cannot avoid potential load shedding, so the capacity of the GB is increased accordingly. By comparing the two cases of $\Gamma^D_n=2$ and $\Gamma^D_n=3$, it is obvious that in the case of the same GB capacity invested, the longer the contingency period is considered, the larger the capacity of the TESS should be built. The same conclusion can also be obtained by comparing the three cases of $\Gamma^D_n=4$, $\Gamma^D_n=5$, and $\Gamma^D_n=6$. It leads to the second function of TESS, improving IES reliability by providing reserve. 

% \begin{table}
%   \centering
%   \caption{The capacity of TESS and GB under different $\Gamma^D_n$.}
%   \begin{tabular}{ccc}
%     \hline
%     \hline
%        $\Gamma^D_n$ & TESS/MWh & GB/MW \\
%     \hline
%     0   & 3.05 & 50 \\
%     1,2 & 69.81 & 10+30\\
%     3   & 89.24 & 10+30\\
%     4   & 33.00 & 10+20+30\\
%     5   & 35.58 & 10+20+30\\
%     6   & 36.01 & 10+20+30\\ 
%     \hline
%     \hline
%   \end{tabular}
%   \label{tab:capacity of TESS and GB}
% \end{table}

\begin{table}
  \centering
  \caption{The capacity of TESS and GB under different $\Gamma^D_n$.}
  \begin{tabular}{ccc|ccc}
    \hline
    \hline
       $\Gamma^D_n$ & TESS/MWh & GB/MW & $\Gamma^D_n$ & TESS/MWh & GB/MW\\
    \hline
    0   & 3.08 & 50 & 4 & 32.83 & 10+20+30\\
    1,2 & 69.81 & 10+30 & 5 & 35.35 & 10+20+30\\
    3   & 89.24 & 10+30 & 6 & 35.78 & 10+20+30\\
    \hline
    \hline
  \end{tabular}
  \label{tab:capacity of TESS and GB}
\end{table}

% \begin{table}
%   \centering
%   \caption{The capacity of TESS and GB under different $\Gamma^D_n$.}
%   \begin{tabular}{ccccccc}
%     \hline
%     \hline
%        $\Gamma^D_n$ & 0 & 1,2 & 3 & 4 & 5 & 6 \\
%     \hline
%     TESS/MWh & 3.05 & 69.81 & 89.24 & 33.00 & 35.58 & 36.01\\
%     GB/MW & 50 & 10+30 & 10+30 & 10+20+30 & 10+20+30 & 10+20+30\\
%     \hline
%     \hline
%   \end{tabular}
%   \label{tab:capacity of TESS and GB}
% \end{table}

\begin{figure*}
  \centering
  \includegraphics[width=0.95\linewidth]{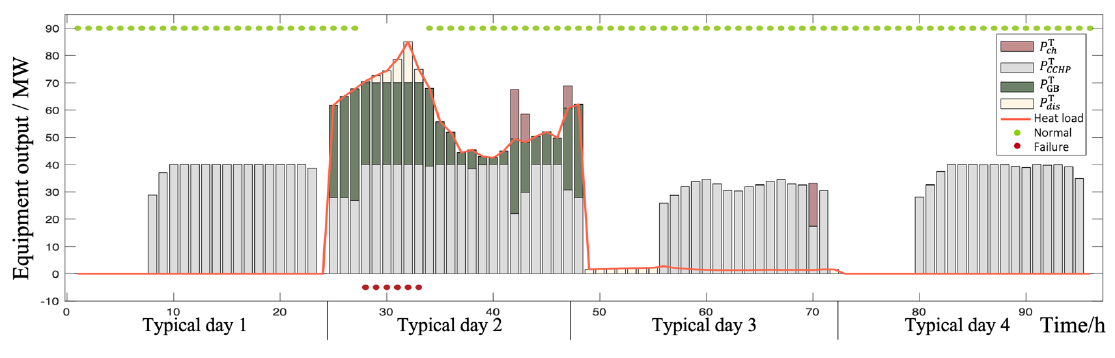}
  \caption{The optimal operation scheme under the planning scheme of $\Gamma^D_n=6$ when the worst-case equipment contingency is considered.}
\label{fig:operation}
\end{figure*}

We choose the case with $\Gamma^{D}_{n}=6$ to further illustrate this function of TESS. Fig. \ref{fig:operation} shows the optimal operation scheme under this planning scheme when the worst-case equipment contingency is considered. It can be seen that under the worst case, the 30MW GB is out of operation while the heat load peaks. We use red dots to indicate the failure state of this GB, while green dots represent its normal operating state. In this scenario, the heat supply by operating boilers and CCHP units, which represented by $P^{\text{T}}_{\text{GB}}$ and $P^{\text{T}}_{\text{CCHP}}$ respectively, can only supply 70 MW of heat, which is below peak demand. However, the stored heat in the TESS $P^{\text{T}}_{dis}$, indicated by the beige bars, compensates for this shortfall, ensuring stable energy supply during equipment contingencies.

%Fig. \ref{fig:operation2} shows the operation scenario that includes the contingency of the 35MW CCHP unit. The contingency happens at almost the same time as the GB contingency that we discussed above to cause a maximum shortage of heat supply. During this CCHP contingency, all the remaining normal equipment can only supply about 70MW of heat power, and the remaining gap is also filled by the heat reserve in advance. In this case, the function of TESS in terms of improving reliability and saving total cost is verified. 

% \begin{figure*}
% \centering
% \subfigure[The operation scheme in case of contingency of the third GB (30MW)]{\includegraphics[width=0.85\linewidth]{figure/operation1.pdf}\label{fig:operation1}}
% \hspace{0.1cm}
% \subfigure[The operation scheme in case of contingency of the second CCHP unit (35MW)]{\includegraphics[width=0.85\linewidth]{figure/operation2.pdf}\label{fig:operation2}}
% \caption{The optimal operation scheme under the planning scheme of $\Gamma^D_n=6$ when the worst-case equipment contingency is considered.}
% \label{fig:operation}
% \end{figure*}

\section{Additional Analysis}
\subsection{Planning Models Comparison}
To better demonstrate the advantages of the proposed two-stage robust planning model, we compare it with other IES planning methods, particularly those focus on reliability. This includes the widely adopted N-1 model \cite{RN171}, N-k-$\epsilon$ model \cite{RN220}, and stochastic optimization (SO) planning models that account for various uncertainties \cite{RN179,RN218}. Specifically, the N-1 model requires the system to continue operating normally in the event of any single equipment failure. The N-k-$\epsilon$ model addresses the overly conservative nature of the N-1 model to some extent by allowing up to k equipment outages with a maximum of $\epsilon$\% load shedding of the total load. Stochastic model considers uncertain load and equipment operational states, deriving the optimal solution based on a large number of scenarios generated by methods like MCS.

The planning results of different planning models are shown in Table \ref{tab:planning result-comparisons}. The reliability and cost comparisons of these planning schemes are shown in Table \ref{tab:reliability and cost-comparisons}. We use total EENS to reflect the reliability here for simplicity.

\begin{table}
  \centering
  \caption{Comparison of planning results under different methods. }
  \begin{tabular}{cccccc}
    \hline
    \hline
        & \begin{tabular}[c]{@{}c@{}}CCHP\\ /MW\end{tabular} & \begin{tabular}[c]{@{}c@{}}GB\\ /MW\end{tabular} & \begin{tabular}[c]{@{}c@{}}EC\\ /MW\end{tabular} & \begin{tabular}[c]{@{}c@{}}BESS\\ /MWh/MW\end{tabular} & \begin{tabular}[c]{@{}c@{}}TESS\\ /MWh\end{tabular} \\
    \hline
    \textit{Deterministic}  & 15+35 & 50 & 22.5 & 7.20/2.13 & 3.08\\
    \textit{N-1-25\%}   & 15+35 & 10+20+30 & 42.5 & 7.20/2.13 & 0\\
    \textit{N-1} & 20+30 & 10+30+40 & 50 & 8.67/2.47 & 9.72\\
    \textit{SO}   & 15+35 & 20+30 & 32.5 & 9.27/3.21 & 2.00\\
    \textit{Proposed}  & 20+30 & 10+20+30 & 55 & 7.80/2.28 & 35.78\\
    \hline
    \hline
  \end{tabular}
  \label{tab:planning result-comparisons}
\end{table}

\begin{table}
  \centering
  \caption{Comparison of EENS and costs of different planning schemes.}
  \begin{tabular}{cccc}
    \hline
    \hline
       & Total EENS/MW & Total cost/¥$10^4$ & Cost increment\\
    \hline
    \textit{Deterministic}   & 95.80 & 241968.1 & - \\
    \textit{N-1-25\%}  & 4.84 & 243259.7 & 0.53\%\\
    \textit{N-1}   & 0.33 & 246707.0 & 1.96\%\\
    \textit{SO}   & 9.98 & 242935.5 & 0.40\%\\
    \textit{Proposed}   & 0.15 & 246159.1 & 1.73\%\\
    \hline
    \hline
  \end{tabular}
  \label{tab:reliability and cost-comparisons}
\end{table}

We can see that the N-1 model is relatively conservative, leading to higher investment costs. Despite this, its reliability is still inferior to the scheme obtained from the proposed model. On the other hand, while the results obtained from the N-k-25\% model are less conservative, its reliability is also significantly compromised. These results can be attributed to the fact that the way the N-1 model considers equipment failures is unreasonable: they assume equipment fails during the whole planning period. Consequently, it overlooks the ability of ESS to support short-term failures and cannot make the optimal choice. 

Although the SO model considers both load fluctuations and equipment operational uncertainties, the reliability of the planning scheme is not high. This is because such methods are highly dependent on the accuracy of the generated scenarios. The MCS method generates load and equipment operational states independently, failing to account for their correlation, which leads to a discrepancy between the generated scenarios and actual operational conditions. In contrast, the proposed RO method identifies the worst-case scenario based on uncertainty sets, requiring only the generation of representative scenarios that include load variations. Moreover, since most equipment failures occur during high load demand, considering uncertainty sets for both equipment failures and load fluctuations simultaneously, as we did in \textit{Case 4}, is reasonable without being overly conservative. 

\subsection{Convergence Performance of Nested C\&CG}
\begin{figure}
  \centering
  \includegraphics[width=0.9\linewidth]{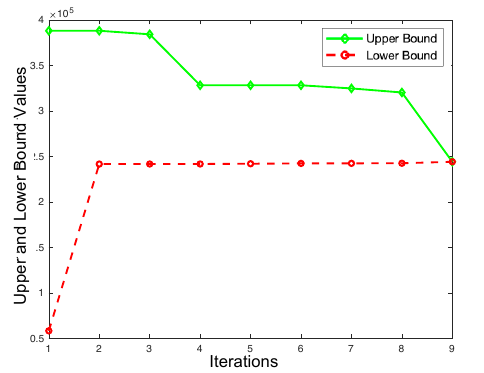}
  \caption{Convergence process of the outer layer algorithm.}
  \label{fig:convergence-outer}
\end{figure}

In this subsection we use two examples to better analyze the solution process and demonstrate the computational efficiency of our reformulated algorithm. First, we chose a case with $\Gamma^{D}_{n}=6$, without considering load uncertainty or substation limits, as the traditional KKT-based nested C\&CG algorithm struggles to handle model involving these factors. Regarding the convergence performance, the outer layer algorithm requires a total of 8 iterations before reaching convergence. The upper and lower bound values are depicted in Fig. \ref{fig:convergence-outer}. 

The master problem optimizes the first stage decision variables and yields a lower bound. This lower bound gradually increases as the planning scheme becomes more reliable. Conversely, the subproblem boosts the total cost by finding the worst operation scenario with equipment contingency. In this process, the load shedding occurs and costs a high penalty and therefore gets an upper bound with a large value. As the planning scheme gradually becomes reliable, there will be no load shedding even in the worst operation scenario and the problem converges to the optimum.

% Based on the principle of the nested C\&CG algorithm, each outer iteration involves multiple iterations of the inner C\&CG algorithm, which are used to solve the subproblem. In our case, all inner subproblems are solved after only 2-3 iterations. The convergence of the inner layer algorithm is very fast, and the gap becomes relatively small after only 1-2 iterations.

\begin{figure}
  \centering
  \includegraphics[width=0.9\linewidth]{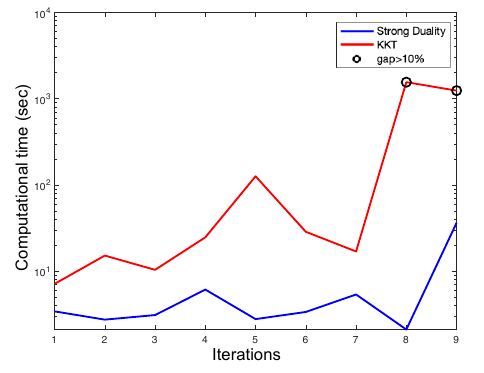}
  \caption{Computational time for subproblem reformulated by KKT and strong duality in each outer layer iteration.}
  \label{fig:convergence-speed}
\end{figure}

% The efficiency of the nested C\&CG when applied to solve the proposed two-stage robust IES planning model will be demonstrated.
In this case, the master problem is less time-consuming compared to the subproblem. This is because the subproblem is reformulated into a tri-level problem, and its solution process is another C\&CG. We use KKT conditions and SD with the linearization method \eqref{general-reform} to reformulate the subproblem respectively. The computational time of the subproblem in each outer iteration is shown in 
Fig. \ref{fig:convergence-speed}. It is obvious that the solution speed of the SD-based subproblem is much faster than that of KKT-based subproblem. Moreover, the KKT-based subproblem still maintains a gap of over 10\% after more than 1000 seconds of computation in the last two iterations. Other iterations eventually converge to a gap of 0.01\%.

We now compare two linearization methods for the SD-based subproblem, i.e., \eqref{general-reform} and the one replaces \eqref{reform} with \eqref{new-reform}. We use the examples considering uncertain loads with 4 and 8 typical days respectively. The total solution time for subproblems is shown in Table \ref{tab:subproblem solution time}. The proposed linearization method effectively reduces the total solution time for subproblem, which is also the main part of the total solving time.

\begin{table}
  \centering
  \caption{Comparison of total solving time for subproblems under different linearization method.}
  \begin{tabular}{cccc}
    \hline
    \hline
       & \textcolor{blue}{\eqref{general-reform}} & Replace \eqref{reform} with \eqref{new-reform} & Time reduction\\
    \hline
    4 typical days  & 10534.0 & 2498.9 & 76.3\% \\
    8 typical days   & 9997.6 & 5556.3 & 44.4\%\\
    \hline
    \hline
  \end{tabular}
  \label{tab:subproblem solution time}
\end{table}

The first example indicates that the SD-based subproblem is more efficient that its KKT-based counterpart. Although both methods introduce bilinear terms into the planning model and use binary variables to eliminate these bilinear terms, the SD-based method results in significantly fewer binary variables than the KKT-based method. Consequently, the computational efficiency of the former is much higher than that of the latter.
Specifically, in our case, KKT introduces over 14000 binary variables, whereas that SD introduces less than 3000 binary variables in each inner layer iteration, which is a huge difference. Additionally, our linearization method resulting in an MILP model, which is more tractable than the MIQCP derived from the typical SD-based nested C\&CG approach. Moreover, by leveraging the mutual exclusivity of the binary variables from uncertain load set, the proposed linearization method further accelerates the solution process as the second example shows. Since in most energy system planning problems, the load uncertainty set can be constructed in this way, the proposed linearization method is applicable across various cases.

\subsection{Uncertainty Budgets and Typical Days Selection}
In the proposed equipment contingency uncertainty set, there are three adjustable uncertainty budgets: $\Gamma^{N}$, $\Gamma^{I}_n$ and $\Gamma^{D}_n$. However, we only analyzed the impact of $\Gamma^{D}_n$ on the planning results in last section. Here, we will explain the effects of the other two budgets on the results. Budget $\Gamma^{N}$ represents the maximum number of simultaneous equipment contingencies we consider. It was fixed at 1 because considering the simultaneous failure of 2 or more pieces of equipment is unreasonable in capacity planning problem for park-level IES. Moreover, since robust optimization is designed to find the solution for the worst-case scenario, considering 2 or more failures would lead to overly conservative results, which makes the planning results meaningless for practical application. Nevertheless, we include this budget in the proposed uncertainty set to allow flexibility for other contexts. For example, in large-scale transmission network expansion planning, where many lines and equipment are involved, the simultaneous failure of 2 or more components becomes a realistic consideration. 

For $\Gamma^{I}_n$, it represents the interval between two consecutive failures, reflecting the frequency of failures. In the previous section, we set it as 48, meaning that up to two failures could occur within the Period=96 we considered. Now we set $\Gamma^{I}_n$=24, 48, and 72 respectively with $\Gamma^{D}_n=6$. The planning results are exactly same in all cases. However, when we changed the values of $\Gamma^{D}_n$ in last section, the results varied significantly. This suggests that prolonged equipment contingencies necessitate capacity expansion to provide more reserve, while the interval between sequential contingencies has relatively less impact on energy supply as long as total capacity remains sufficient. 

One of the key advantages of our proposed uncertainty set is that its budget can directly correspond to commonly used reliability metrics in most studies on reliability. For example, $\Gamma^{I}_n$ corresponds to equipment failure rate, and $\Gamma^{D}_n$ relates to equipment repair rate or repair/failure duration. From \cite{RN203}, \cite{RN344}, we know the failure rates of the considered equipment are around 1 to 5 times per year and the repair rate is around 0.15 to 0.2. From \cite{RN330}, the failure rate of distribution line is 0.096 failures per kilometer per year, and the repair time is around 4 hours. Based on this, setting $\Gamma^{I}_n$ to 24, 48, and 72, and $\Gamma^{D}_n$ from 1 to 6, as done in the previous sections, is reasonable.

As mentioned earlier, we aim to discuss whether it is reasonable to use four typical days as the basis for planning and to incorporate these reliability parameters into the typical days in this manner. An obvious limitation of using four typical days is that when considering a failure frequency greater than 4, we would need to account for the occurrence of two failures on the same day, which is uncommon in practice. Therefore, we employ K-means to generate eight typical days and compare the planning results with \textit{Case 3} and \textit{Case 4}. We set $\Gamma^{D}_n=6, \Gamma^{I}_n=30$. When the load uncertainty is not considered, the result is the same as that of \textit{Case 3}. And when we consider the uncertain load, the results showed only a minor difference from \textit{Case 4}, with a reduction of 0.79 MWh in BESS investment. The results indicate that the four selected typical days are sufficiently representative and the selection of uncertainty set budgets are reasonable.

\section{Conclusion}
Reliability is one of the most important performance indicators for IES, as it directly affects energy supply and is closely related to the functioning of society and people's lives. This paper proposes a two-stage robust planning model for park-level IES, which can avoid the failure of the system under the interference of uncertainty such as load fluctuation and equipment contingency. The conclusions drawn from this study are threefold: 
\begin{itemize}
  \item[1)] The proposed robust IES planning model outperforms other widely used models in improving system reliability. In particular, the robust planning scheme that considers short-term equipment contingency significantly reduces EENS, LOLE, and LOLF, by over 90\% compared to the deterministic planning scheme, with a marginal increase in total cost of only 1.53\%. The most robust scheme considering longer contingency and load uncertainty nearly eliminates these metrics and increases the total cost by around 2\%;
  \item[2)] The ESS plays an important role in IES. In this study, the roles of both BESS and TESS in reducing operational costs and enhancing system reliability are analyzed. Additionally, the power-to-energy ratio of the BESS can be optimized to address contingencies of different durations;
  \item[3)] The nested C\&CG algorithm performs well in solving two-stage robust IES planning problems with second-stage integer variables. In the case study, we showed that the SD-based nested C\&CG algorithm is more efficient than the KKT-based approach, with further improvements from our proposed linearization method. 
\end{itemize}

This work provides a reference for park-level IES planning, and the proposed method is versatile, as other types of equipment or different uncertainty sets can be easily integrated into this framework. One straightforward work for future is to extend this planning approach to interconnected multi-regional IESs. This would require consideration of more components, their operational states, and various other uncertainties, introducing a significant number of variables. Such an extension would pose substantial challenges to the scalability of the model. However, the scalability of the model is inherently limited by its formulation as an MILP, as solving large-scale MILP problems remains challenging. Therefore, our future work will focus on developing more concise yet accurate modeling approaches, as well as advancing algorithmic acceleration strategies.
% The proposed model will be applied to more complex and larger energy systems, and the proposed uncertainty set will be used to describe all kinds of possible contingencies, e.g., transmission line outage. As the size of the problem increases, the corresponding algorithm should also be further developed to improve the solving speed. 

% or


\appendix
\renewcommand{\theequation}{A.\arabic{equation}}
\setcounter{equation}{0} % 重置编号
\numberwithin{equation}{section}
\section{The Constraints of Reformulated Planning Model}
Here we illustrate how to reformulate \eqref{<theta>} using KKT conditions. The master problem of the subproblem can be shown as:
\begin{subequations}
  \begin{align}
  &\textbf{MPs:   } \max \sigma, \\
    &s.t.\quad \sigma \leq m\sum_{t}[P^{\text{SUB}}_{0,t,r}\mathcal{R}^{e}_{t}+(\sum_i P^{\text{CCHP}}_{0,i,t,r}+\sum_j P^{\text{GB}}_{0,j,t,r}\notag \\&)\mathcal{R}^{g}_{t}+P^{\text{LS},d}_{1,t,r}p^{\text{LS},d}_{t}],\label{<Cons_MPs_begin_kkt>}\\
&\textbf{L} \leq \textbf{C}^{\text{SUB}}\textbf{P}^{\text{SUB}}_{0,r}+\sum_{\text{Equi}}\textbf{C}^{\text{Equi}}\textbf{P}^{\text{Equi}}_{0,r}-\textbf{P}^{\text{ESS}}_{0,r} \label{<Cons_MPs_energy_balance_kkt>}\\
    &\textbf{L} \leq \textbf{C}^{\text{SUB}}\textbf{P}^{\text{SUB}}_{1,r}+\sum_{\text{Equi}}\textbf{C}^{\text{Equi}}\textbf{P}^{\text{Equi}}_{1,r}-\textbf{P}^{\text{ESS}}_{1,r}+\textbf{P}^{\text{LS}}_{1,r}\\
&0 \leq P^{\text{Equi}}_{0,n,t,r}
    \leq P^{\text{Equi}}_{n,max}\hat{X}^{\text{Equi}}_{n}, \\
&0 \leq P^{\text{Equi}}_{1,n,t,r}
    \leq P^{\text{Equi}}_{n,max}s^{\text{Equi}}_{n,t}\hat{X}^{\text{Equi}}_{n},\\
    & 0 \leq P^{\text{SUB}}_{0,t,r} \leq \sum_f P^{\text{SUB}}_{f,max}, \\
    & 0 \leq P^{\text{SUB}}_{1,t,r} \leq \sum_f P^{\text{SUB}}_{f,max}\hat{s}^{\text{feed}}_{f,t},\\
    &Cons^{\text{ESS}}_{0,r}(\hat{X}^{\text{B/T}}_E,\hat{X}^{\text{B/T}}_P,P^{\text{B/T}}_{0,ch/dis,t,r},\tilde{Z}^{\text{B/T}}_{0,ch,t,r},E^{\text{B/T}}_{0,t,r}),\\
& Cons^{\text{ESS}}_{1,r}(\hat{X}^{\text{B/T}}_E,\hat{X}^{\text{B/T}}_P,P^{\text{B/T}}_{1,ch/dis,t,r},\tilde{Z}^{\text{B/T}}_{1,ch,t,r},E^{\text{B/T}}_{1,t,r}),\label{<Cons_MPs_Cons_ESS_kkt>}\\ 
&0 \leq P^{\text{SUB}}_{0/1,t,r}\label{eq:con-MPs-sub_kkt}\\
&0 \leq P^{\text{LS},d}_{1,t,r}, \label{<Cons_MPs_Cons_LS_kkt>}\\
&\text{All dual variables}  \geq 0,\label{eq:MPs-dual_kkt}\\
    &Cons^{La}_r,Cons^{S}_r,\label{eq:MPs-KKT_kkt}\\  
     &l \in{\mathbb{L}}, s \in{\mathbb{S}},\\
    &\forall 1 \leq r \leq p_s .\notag
  \end{align}
\end{subequations}
Constraints \eqref{<Cons_MPs_energy_balance_kkt>}—\eqref{<Cons_MPs_Cons_LS_kkt>} are in similar form as those in the subproblem. Constraint (\ref{eq:MPs-dual_kkt}) is for dual variables corresponding to constraints in primal problem. In addition to the constraints of the primal problem and the constraints for dual variables, KKT conditions also contain other two groups of constraints, namely constraints for the gradient of Lagrange and the complementary slackness conditions. Here we use $Cons^{La}$ and $Cons^{S}$ to denote these two groups of constraints respectively as shown in (\ref{eq:MPs-KKT_kkt}).

We would like to illustrate how to derive $Cons^{La}$ and $Cons^S$ instead of listing all of these constraints here. First of all, ($P^{\text{SUB}}_{0/1,t,r}$, $P^{\text{CCHP}}_{0/1,t,r}$, $P^{\text{EC}}_{0/1,t,r}$, $E^{\text{B/T}}_{0/1,t,r}$, $P^{\text{B/T}}_{0/1,ch/dis,t,r}$, $P^{\text{LS},d}_{1,t,r}$) is set as a group of variables of the primal problem. Then we can derive constraints by making the gradient of Lagrange with respect to each variable equal to 0. Take $P^{\text{SUB}}_{0,t,r}$ as an example, we get:
\begin{equation}
    m\mathcal{R}^e_t-C^{\text{SUB}}_{e-e}\alpha_{0,t,r}-\rho_{0,t,r} = 0, \label{eq:con-la-example_kkt}
\end{equation}
where $\alpha_{0,t,r}$ and $\rho_{0,t,r}$ are dual variables for constraints $l^e_t \le C^{\text{SUB}}_{e-e}P^{\text{SUB}}_{0,t,r}+\sum_{i}C^{\text{CCHP}}_{i,g-e}P^{\text{CCHP}}_{0,i,t,r}+
    \sum_{k}C^{\text{EC}}_{k,e-e}P^{\text{EC}}_{0,k,t,r}
    -P^{\text{B}}_{0,ch,t,r}+P^{\text{B}}_{0,dis,t,r}$ and $P^{\text{SUB}}_{0,t,r} \geq 0$ respectively. We have:
$\alpha_{0,t,r}, \rho_{0,t,r} \geq 0$. It is obvious that we can eliminate $\rho_{0,t,r}$ and transform constraint (\ref{eq:con-la-example_kkt}) into:
\begin{equation}
    m\mathcal{R}^e_t-C^{\text{SUB}}_{e-e}\alpha_{0,t,r} \geq 0.
\end{equation}
In this way we can get all constraints of $Cons^{La}$. As for complementary slackness condition, the first constraint can be list as:
\begin{gather}
    (-l^e_t + C^{\text{SUB}}_{e-e}P^{\text{SUB}}_{0,t,r}+\sum_{i}C^{\text{CCHP}}_{i,g-e}P^{\text{CCHP}}_{0,i,t,r}+
    \sum_{k}C^{\text{EC}}_{k,e-e}\notag \\P^{\text{EC}}_{0,k,t,r}
    -P^{\text{B}}_{0,ch,t,r}+P^{\text{B}}_{0,dis,t,r})\alpha_{0,t,r} = 0. \label{<example_kkt>}
\end{gather}
This is a non-linear constraint which can be reformulated to a linear constraint by using big-M method and introducing binary variables $\nu_{t,r}$:
\begin{align}
    &-l^e_t + C^{\text{SUB}}_{e-e}P^{\text{SUB}}_{0,t,r}+C^{\text{CCHP}}_{i,g-e}P^{\text{CCHP}}_{0,i,t,r}+C^{\text{EC}}_{k,e-e}P^{\text{EC}}_{0,k,t,r} \notag \\
    &-P^{\text{B}}_{0,ch,t,r}+P^{\text{B}}_{0,dis,t,r} \leq M\nu_{t,r},\\
    &\alpha_{0,t,r} \leq M (1-\nu^{}_{t,r}),\\
    &\nu_{t,r} \in \{0,1\}. \notag
\end{align}
$M$ is a relatively big constant. It can be substituted by the upper bound of the left-hand side of the inequality. Then we get $Cons^S$.

The \textbf{SPs} and other steps are the same as that of strong duality-based reformulation.

\section*{Acknowledgment}

The authors would like to acknowledge the help of Prof. Line Roald, both in discussions leading to the development of the paper as well as feedback on the paper draft.
% \section*{Acknowledgment}


\ifCLASSOPTIONcaptionsoff
  \newpage
\fi

\bibliographystyle{IEEEtran}
\small\bibliography{main}

\end{document}